\documentclass[
twocolumn,
superscriptaddress,
 amsmath, amssymb,
aps,
prl,
dblfloatfix,
nobalancelastpage
]{revtex4-1}
\usepackage{graphicx}
\usepackage{dcolumn}
\usepackage{bm}
\usepackage{siunitx}
\usepackage{hyperref}
\usepackage{braket}
\usepackage{epstopdf}
\DeclareGraphicsExtensions{.eps,.pdf}
\usepackage{filecontents}
\hypersetup{colorlinks=true, citecolor=blue, urlcolor=blue, linkcolor=blue}

\graphicspath{{C:/Users/minwo/Desktop/Course Works/Ultracold term paper}}

\begin{document}

\preprint{APS/123-QED}

\title{Active Valley-topological Plasmonic Crystal in Metagate-tuned Graphene}

\author{Minwoo Jung}
\email{mj397@cornell.edu}
\affiliation{Department of Physics, Cornell University, Ithaca, New York, 14853, USA}
\author{Zhiyuan Fan}
\author{Gennady Shvets}
\email{gshvets@cornell.edu}
\affiliation{School of Applied and Engineering Physics, Cornell University, Ithaca, New York 14853, USA}

\date{\today}

\begin{abstract}
A valley plasmonic crystal for graphene surface plasmons (GSPs) is proposed. We demonstrate that a designer metagate, placed within a few nanometers from graphene, can be used to impose a triangular periodic Fermi energy landscape on the latter. For specific metagate geometries and bias voltages, the combined metagate-graphene structure is shown to produce sufficiently strong Bragg scattering of GSPs to produce complete propagation bandgaps, and to impart the GSPs with nontrivial valley-linked topological properties. Valley-selective kink states supported by a domain wall between differently patterned metagates are shown to propagate without reflections along sharply curved interfaces owing to suppressed inter-valley scattering. Our approach paves the way for non-magnetic dynamically reconfigurable topological nanophotonic devices.
\end{abstract}

\maketitle
Graphene is a promising platform for nanoscale terahertz photonic devices because it supports deeply subwavelength graphene surface plasmons (GSPs) \cite{hBN1}. The propagation of GSPs can be dynamically controlled by injecting free carriers (electrons or holes) using field-effect electrostatic gating~\cite{CD1, CD2, CD3}. Moreover, the electron scattering times in high-quality graphene have been extended to nearly a picosecond, thus enabling extremely long propagation distances of the GSPs~\cite{SG1, SG2, hBN2, hBN3}. Recently reported demonstration of electrical control of the GSP's phase~\cite{hBN4} suggests that graphene-based nanophotonic devices can be now envisioned. The natural next step in developing such devices is to investigate the possibility of topological protection of GSPs, similarly to the way it has been done in microwave and micron-scale photonic structures~\cite{soljac_nature09,hafizi_nature11,shvets_nature_mat_13,segev_nature13,ctchan_nature_comm_14,shvets_prl_15}.

\begin{figure}[t]
    \includegraphics[width=0.95\columnwidth]{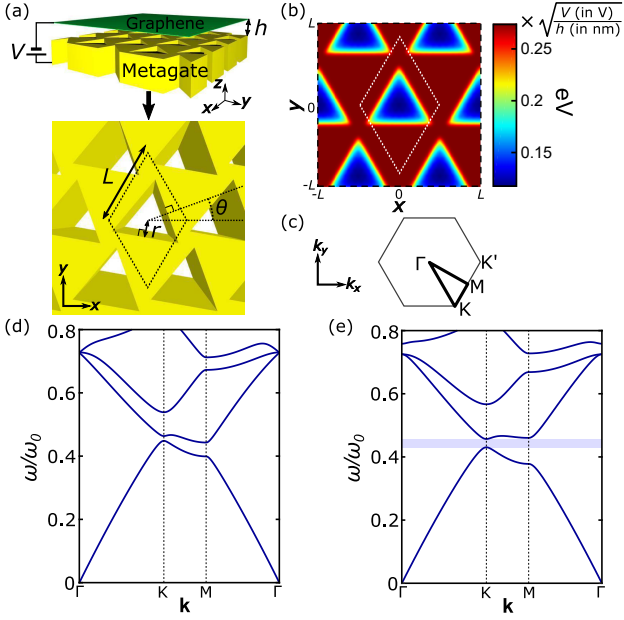}
    \caption{(a) Schematic of a GSP-based VC: a patterned metagate biased by a static voltage $V$ with respect to graphene imprints (b) a periodic landscape of chemical potential $E_F(x,y)$ onto graphene. (c) Brillouin zone of a triangular lattice and its high symmetry points. (d) Dispersion relation of the GSPs propagating through graphene with homogeneous doping ($V=0$, but assuming $E_F(\bold{r})=E_0\neq0$). (e) Same as (d), but for a landscaped chemical potential from (b); the complete bandgap is shaded. Geometric parameters: $\theta = 30^{\circ}$, $L/r = 4$, and $L/h = 25$.}
    \label{fig.1.}
\end{figure}

Several recent proposals~\cite{TGP, TGP2} utilize large magnetic fields to produce topological GSPs by breaking the time-reversal symmetry in nanopatterned graphene. Among the practical limitations of this approach are the lack of reconfigurability (large magnetic fields cannot be rapidly turned off) and the degradation of graphene's quality arising from its patterning\cite{TGP2, Edge}. In this Letter, we propose an alternative approach that requires neither graphene patterning nor magnetic field. Instead, we demonstrate that a regular landscape of chemical potential can be "imprinted" into graphene by placing it in close proximity of an electrically-biased patterned metagate shown in Fig.~\ref{fig.1.}(a). Because graphene's optical conductivity is related to its chemical potential, such landscapes present periodic patterns of the effective refractive index to the GSPs, causing their Bragg scattering and producing a complex propagation band structure. When the metagate pattern is a two-dimensional (2D) triangular crystal with broken mirror symmetry, topological properties linked to the valley degree of freedom emerge in GSPs.

Historically, such valley crystals (VC) were originally discovered in electronic systems and studied in the context of valley-polarized non-zero Hall conductivities \cite{VHI2007, IVS2006, VF2007,  VP2006, VHE2014}, giving rise to the emerging field of valleytronics~\cite{niu_prb08,geim_science14}. More recently, the study of valley-based topological phases was extended to bosonic systems, in which the particular interest is the emergence of confined chiral kink states (CKSs) at the domain wall between two VCs that are mirror images of each other. The propagation of the CKSs is valley-locked and topologically protected in the absence of inter-valley scattering \cite{TCSC2014, TCBLG2008}. Such topologically robust transport of valley-locked CKSs has been extensively studied in several bosonic---photonic/plasmonic \cite{AllSi2016, SV2017, QVH2017, QVH2017-2, Photon2017, wen_natcomm2017} and phononic \cite{Phonon2016, Phonon2017}---systems as well as in electronic systems \cite{TVPBLG2011, TVPBLG2015, TVPBLG2015-2, TVPBLG2016, VPC2016, VPC2012}.

In this Letter, we show that a monolayer graphene electrostatically doped by a metagate provides a nanoscale plasmonic platform for mid-infrared valleytronics. The specific metagate design shown in Fig.~\ref{fig.1.}(a) consists of a half-infinite perfectly conducting metal penetrated by a triangular lattice of infinitely deep equilateral triangular holes (corrections due to finite resolution in patterning and finite depths of the holes are discussed in Supplemental Material \cite{Supp}). When a static voltage $V=V_{\text{MG}}-V_{\text{gr}}$ is applied between the metagate (at an electric potential $V_{\text{MG}}$) and graphene (at an electrochemical potential $V_{\text{gr}}$: $-eV_{\text{gr}} = E_F(\bold{r}) - e\phi_{\text{gr}}(\bold{r})$~\cite{novoselov_nnano08}), the latter is electrostatically doped by free electrons with a spatially-dependent surface density $n(\bold{r})$, where $E_{F}(\bold{r}) = \hbar v_{F} \sqrt{\pi n(\bold{r})}$($v_F \sim 10^6$m/s is the Fermi velocity) is the chemical potential. Here $\bold{r} \equiv (x,y)$ and the graphene sheet is located at $z=0$. The $v_F-$renormalization from electron-electron interactions~\cite{novoselov_pnas13, koppens_science_17} is ignored for the discussions below because its influence on the GSP dispersion was calculated to be negligible, see Supplemental Material \cite{Supp}. An example of the chemical potential landscape $E_F(\bold{r})$ in graphene is shown in Fig.~\ref{fig.1.}(b) for the geometric parameters of the metagate listed in the caption. Clearly, the metagate imprints a VC onto graphene with the spatial symmetry ($C_3$) required for realizing a valley-Hall photonic topological insulator~\cite{AllSi2016}.

The static electron density $n(\bold{r})$ is determined from $n(\bold{r}) = (\epsilon_0/e)\left[\partial_z \phi(\bold{r},z)|_{z=-0} - \partial_z\phi(\bold{r},z)|_{z=+0}\right]$, where the static potential $\phi(\bold{r},z)$ is a solution to the Laplace equation $\nabla^2 \phi =0$ with self-consistent boundary conditions: $\phi(\bold{r},z=0) = \phi_{\text{gr}}(\bold{r}) = V_{\text{gr}} + \frac{\hbar v_{F}}{e} \sqrt{\pi n(\bold{r})}$ and $\phi(\text{MG}) = V_{\text{MG}}$. Due to graphene quantum capacitance (GQC) \cite{QCG}, the boundary conditions themselves are dependent on the final solution $n(\bold{r})$, thus requiring an iterative method to numerically solve the Laplace problem (see Supplemental Materials \cite{Supp}). The GQC effect can be neglected in the limit of $\rho_1 \equiv l_s/h \ll 1$ (equivalently, $E_F(\bold{r}) \ll e|V|$)~\cite{novoselov_nnano08,novoselov_pnas13}, where $l_s=\frac{\pi\epsilon_s\hbar^2 v_F^2}{e^2 E_{F_0}}$ is the graphene static screening length~\cite{San} ($\epsilon_s$ is the static permittivity of the spacer material, and $E_{F_0}=\langle E_F(\bold{r})\rangle$ is the averaged value over a unit cell). We note that the normalized screening length scale $\rho_1$ also determines the velocity of the acoustic GSPs $v_{\rm ac} \equiv \omega_{\rm ac}({\bf q})/|{\bf q}|$ according to $v_{\rm ac}/v_F = (1 + \rho_1/2)/(\rho_1 + \rho_1^2/4)^{1/2} > 1$~\cite{hBN3, koppens_science_17}.  With GQC ignored, graphene in the static limit can be treated as a classical conducting sheet ($\phi_{\text{gr}}(\bold{r}) = V_{\text{gr}}$), and the solution $n(\bold{r})$ of the Laplace equation (subject to $n(\bold{r})$-independent boundary conditions) is scale-invariant, i.e. the normalized chemical potential profile $\theta(\bold{r}/L) \equiv E_F(\bold{r}/L)/\sqrt{V/h}$ depends only on the geometric proportions (e.g., $L:r:h$) of the structure.

\begin{figure}[b]
    \includegraphics[width=0.92\columnwidth]{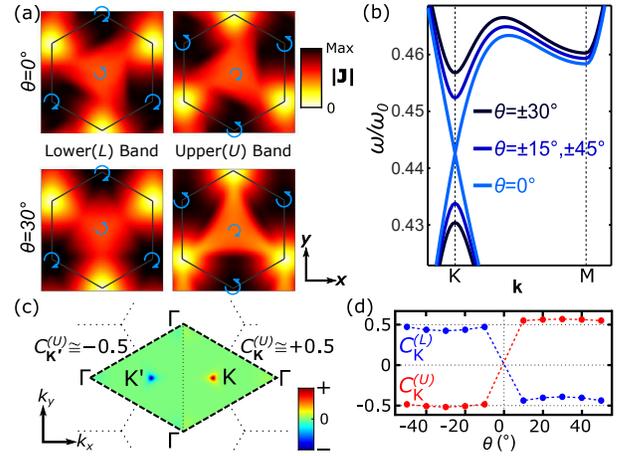}
    \caption{(a) Current density profiles ${\bf J} = \sigma {\bf \nabla} \delta \phi$ of GSPs at $\bold{K}$ (or $\bold{K'}$) on the lower and the upper bands respect to the bandgap; the color images are time-averaged magnitudes of ${\bf J}$, and the blue arrows indicate the rotation direction of local ${\bf J}$ vectors. (b) Bandgap at the Dirac points. (c) Composite Berry flux $F^{(U)}(\bold{k})$ of the upper bands ($2\oplus3$) in $\theta = +30^{\circ}$ case, and the associated valley Chern numbers. (d) Valley chern numbers at several values of $\theta$.}
    \label{fig.2.}
\end{figure}

\begin{figure}
    \includegraphics[width=0.92\columnwidth]{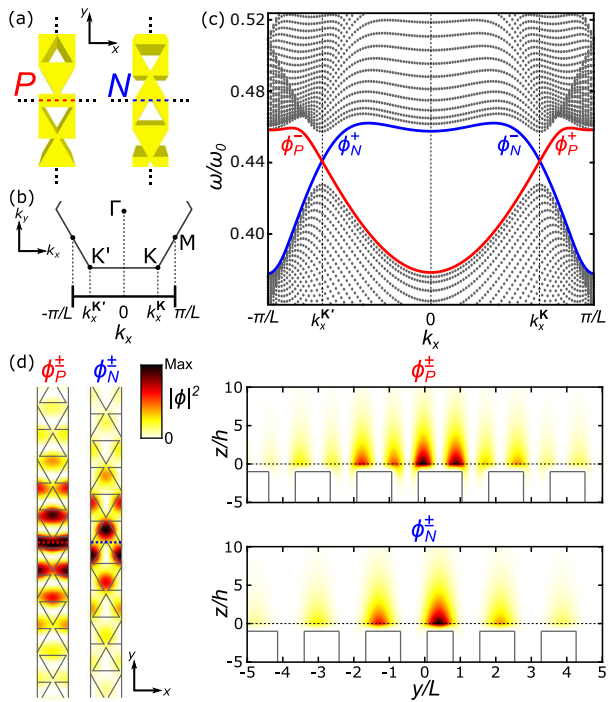}
    \caption{Valley-selective CKSs at interfaces of two domains, $\theta = \pm30^{\circ}$. (a) Two types of domain walls. (b) Projected 1D BZ (c) 1D edge dispersion and topological CKSs; CKSs are labeled by the type of supported domain walls (subscripts) and the direction of the group velocity (superscripts). (d) Left: Potential profiles of CKSs on graphene plane. Right: Potential profiles of CKSs on a $y-z$ cut-plane ($x=0$).}
    \label{fig.3.}
\end{figure}

The band dispersion of GSPs is obtained from the coupled integral equations for the optical perturbations of the potential $\delta \tilde{\phi}$ and charge density $\delta \tilde{n}$~\cite{Book1}:
\begin{subequations}
\label{eq1}
\begin{gather}
\delta \tilde{\phi}(\bold{q}, \omega) = \sum\limits_{\bold{q'}} \tilde{v}\left(\bold{q},\bold{q'},\omega\right) \delta \tilde{n}(\bold{q'}, \omega) \label{eq1a}\\
\delta \tilde{n}(\bold{q}, \omega) = \sum\limits_{\bold{q'}} \tilde{\chi}_{nn}\left(\bold{q},\bold{q'},\omega\right) \delta \tilde{\phi}(\bold{q'}, \omega) \label{eq1b}
\end{gather}
\end{subequations}
where the quantities under a tilde are calculated in the momentum/frequency Fourier domains. Equation (\ref{eq1a}) represents the Poisson's equation for 2D charge distributions, where $\tilde{v}$ describes the 2D screened Coulomb interaction in the presence of surrounding dielectrics and/or metal metagates. While $\tilde{v} \propto \delta_{\bold{q},\bold{q'}}$ for an un-patterned gate, the non-trivial $(\bold{q},\bold{q'})$-dependence of $\tilde{v}$ arises as the periodic nature of the metagate induces a periodic array of image charges. Thus, a density perturbation with the wavevector $\bold{q}$ induces harmonic overtones of the potential with a set of wavevectors $\bold{q'} = \bold{q} + \{\bold{G}\}$ ($\{\bold{G}\}$ is the set of the reciprocal lattice vectors). An exact analytic form of $\tilde{v}\left(\bold{q},\bold{q'},\omega\right)$ in the presence of a periodic metagate is derived in the Supplemental Material \cite{Supp}. Equation (\ref{eq1b}) expresses the charge continuity, and contains the density response function $\tilde{\chi}_{nn}$ that is related to the non-local conductivity $\tilde{\sigma}$ according to $\tilde{\chi}_{nn}\left(\bold{q},\bold{q'},\omega\right) = \frac{\bold{q}\cdot\bold{q'}}{i\omega e^2} \tilde{\sigma}\left(\bold{q},\bold{q'},\omega\right)$ (see Supplemental Material \cite{Supp} for a more detailed derivation). For the simplest example of graphene suspended in vacuum (no metagate) and doped homogeneously ($E_F({\bold{r}}) = E_{F_0}$), we have $\tilde{v} = \delta_{\bold{q},\bold{q'}} \frac{e^2}{2\epsilon_0|\bold{q}|}$ and $\tilde{\chi}_{nn} = \delta_{\bold{q},\bold{q'}} \frac{E_{F_{0}}\bold{q}\cdot\bold{q'}}{\pi \hbar^2\omega^2}$ in local Drude approximation~\cite{Drude}. Substituting these expressions into Eqs.~(\ref{eq1}) gives the well-known dispersion relation for GSPs in suspended graphene\cite{San}: $\omega_{\rm susp}\left(\bold{q}\right) = \sqrt{e^2 E_{F_{0}}|\bold{q}|/2\pi\epsilon_0\hbar^2}$.

The nontrivial $(\bold{q},\bold{q'})$-dependence of $\tilde{\sigma}$ originates from two sources: (i) the non-uniform spatial distribution of the unperturbed static electron density $n(\bold{r})$, and (ii) the inherently non-local electron response (NLER) in graphene \cite{chinn} that grows as fourth-order in $\rho_2 = |{\bf q}| l_{\rm NL}$, where $l_{\rm NL} = v_F/\omega$ is a non-local Thomas-Fermi screening length~\cite{smith_pendry,smith_pendry_science12}. Note that while standard commercial finite-element electromagnetics codes (e.g., COMSOL) can take into account (i) by assuming the standard (local) Drude conductivity, they cannot account for (ii). Therefore, even though Eqs.(\ref{eq1}) are limited to the relevant electrostatic limit~\cite{koppens_NL_11}, they can model additional, and potentially more important, physical effects that are described later in this Letter.

When GQC and NLER are negligible in $\rho_1,\rho_2 \ll 1$ limits, a scale-invariant $E_F(\bold{r})$ and a local Drude conductivity $\sigma\left(\bold{r},\bold{r'},\omega\right) = \frac{ie^2 E_{F}\left(\bold{r}\right)\delta(\bold{r}-\bold{r'})}{\pi \hbar^2\omega}$ (equivalently, $\tilde{\chi}_{nn}(\bold{q},\bold{q'},\omega) = \frac{\tilde{E}_{F} (\bold{q}-\bold{q'})\bold{q}\cdot\bold{q'}}{\pi \hbar^2\omega^2}$) can be assumed. Then, if the permittivity dispersion of the medium encapsulating graphene is negligible as well, Equations~(\ref{eq1}) can be recast as a linear eigenvalue matrix problem, see Supplemental Material \cite{Supp}, and the resulting plasmonic dispersions are also scale-invariant with the natural frequency scale given by $\omega_{0} = \omega_{\rm susp}\left(|\bold{q}|=\frac{2\pi}{L}\right)$. With $L=\SI{200}{\nano\meter}$ and $E_{F_0} = 0.2$eV, for example, we get $\omega_0/2\pi= 33$THz$\sim 0.02 c/L$, which confirms that our model lies well in the electrostatic limit. A similar model was used \cite{Abajo} in the absence of a patterned metagate.

On the other hand, if GQC, NLER or the permittivity dispersion of the spacer medium cannot be ignored, the plasmonic dispersions are no longer scalable as the operator of the eigenvalue problem itself acquires a frequency dependence. For simplicity in demonstrating generic concepts of GSP-based valley-topological transport, we start by working in a vacuum and neglecting those non-scalable effects, which is actually valid for experimentally reasonable system sizes and bias field gradients ($L > \SI{100}{\nano\meter}$ and $V/h > 0.5$V/nm, see Supplemental Material \cite{Supp}). The example corresponding to significant GQC and NLER corrections in the presence of dispersive graphene-encapsulating materials (e.g. hBN) is discussed later. We neglect the hybridization of the metagate-supported spoof surface plasmons (SSP) \cite{SSP} with GSPs because the coupling between SSPs and GSPs is negligible\cite{SSPgraphene} in the strongly-electrostatic limit.

Before examining the effect of $E_F(\bold{r})$ landscaping, we first consider a situation where GSPs propagate over the metagate in graphene homogeneously doped by other means (i.e. $V=0$ in Fig.~\ref{fig.1.}(a), but $E_F(\bold{r}) = E_{F0} \neq 0$). A propagation bandgap opens at $\bold{K}$ point of the Brillouin zone (BZ) due to the broken mirror symmetry in the metagate structure as observed in Figure.~\ref{fig.1.}(d). The overall effect of the metagate screening and the emergence of the acoustic GSP~\cite{hBN3, AP} is also apparent from the linear $\omega = v_{\rm ac} |{\bf q}|$ dispersion near $\bold{\Gamma}$-point. The periodic screening from the metagate itself is, however, insufficient for opening a complete bandgap despite its proximity($L/h=25$) to graphene. The complete bandgap opens over the whole BZ, as shown in Fig.~\ref{fig.1.}(e), only when a periodic $E_F(\bold{r})$ landscape (Fig.~\ref{fig.1.}(b)) is introduced by biasing($V\neq0$) metagate respect to graphene.

The magnitude of bandgap depends on the orientation angle $\theta$ of the triangular holes defined in Fig.~\ref{fig.1.}(a). For $\theta \equiv 0 ({\rm mod} 60^{\circ})$, the system is mirror-symmetric with respect to $\bold{K}$ and $\bold{K'}$ directions. Therefore, the spatial profiles of the two lowest energy eigenstates at each valley are mirror images of each other (see Fig.~\ref{fig.2.}(a)), thus forming degenerate Dirac points~\cite{SV2017}. In contrast, these two states are no longer degenerate (see Fig.~\ref{fig.2.}(b)) when the mirror symmetry is broken for $\theta \not\equiv 0 ({\rm mod} 60^{\circ})$, thereby producing valley-topological bandgaps.

Topological properties of our GSP-based VC are investigated first through the valley Chern numbers, $C^{(L,U)}_{\nu} \equiv (2\pi i)^{-1}\int_{\bigtriangleup^{\bold{\Gamma}}_{\nu}}F^{(L,U)}(\bold{k}) d^{2}\bold{k}$, evaluated numerically from the computed eigenstates $\delta \tilde{\phi}^{(L,U)}$ \cite{Fukui}. Here, $\nu$ is the valley index ($\bold{K}$ or $\bold{K'}$), $L$ or $U$ corresponds to the lower($L$) and upper($U$) bands with respect to the bandgap, $F^{(L,U)}(\bold{k})\equiv \nabla_{\bold{k}} \times \Bra{\delta \tilde{\phi}_{n\bold{k}}}\nabla_{\bold{k}}\Ket{\delta \tilde{\phi}_{n\bold{k}}}$ is the Berry curvature, and $\bigtriangleup^{\bold{\Gamma}}_{\nu}$ is a triangle defined by the nearest $\bold{\Gamma}$ points from $\nu$. As expected, the valley Chern numbers are close to $\pm 0.5$ \cite{VHI2007}, and their signs are reversed as the binary valley index is changed from $\bold{K}$ to $\bold{K^{\prime}}$, or as $\theta$ mod $60^{\circ}$ crosses $0^{\circ}$, see Fig.~\ref{fig.2.}(c)-(d). Small deviations of the valley Chern numbers from $\pm0.5$ are due to the contribution near $\Gamma$ points, which is not relevant to valley-based dynamics.

When two domains with opposite signs of $\theta$ are interfaced, the difference between the valley Chern numbers from each domain, $\Delta C_{\nu}$, is $\pm1$, resulting in a single CKS at each valley~\cite{macdonald_prb_2011}. The sign of $\Delta C_{\nu}^{(U)}$ indicates the propagation direction of CKSs. As illustrated in Fig.~\ref{fig.3.}(a), there are two types of domain walls, which we label as P(positive)-type or N(negative)-type. At the P-type domain wall, $\theta=30^{\circ}$ domain is placed above $\theta=-30^{\circ}$ domain. Because $\Delta C_{\bold{K}}^{(U)}=+1$ and $\Delta C_{\bold{K'}}^{(U)}=-1$, the P-type domain wall supports a forward-(backward-) propagating CKS in the $\bold{K}$($\bold{K'}$) valley. The situation is reversed at the N-type domain wall, where the $\theta=30^{\circ}$ domain is placed below $\theta=-30^{\circ}$ domain.

\begin{figure}
    \includegraphics[width=0.92\columnwidth]{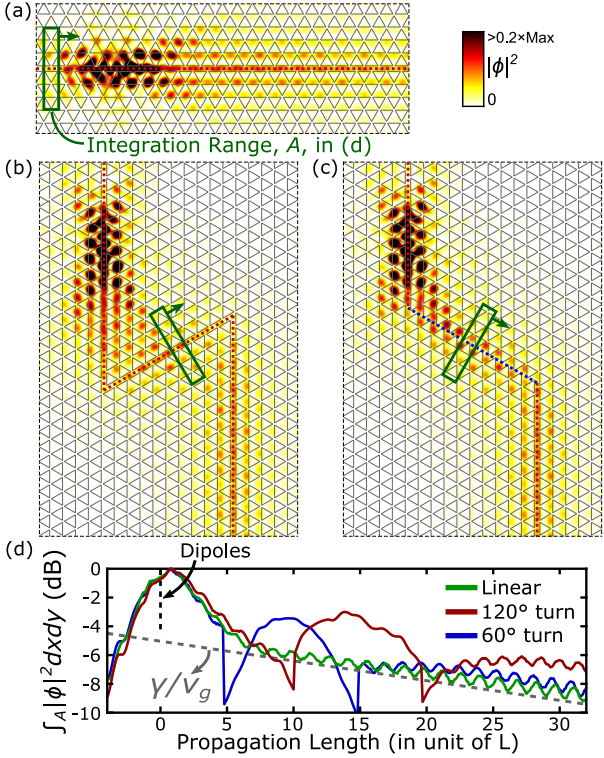}
    \caption{Topological valley transport of GSPs ($\gamma=\omega_{\text{mid}}/300$). (a) Linear Propagation of $\phi_P^{+}$ state. (b) Reflection-free $120^{\circ}$ turn. (c) Reflection-free $60^{\circ}$ turn. (d) Attenuation of the CKSs in (a)-(c) ; the gray dotted line shows the attenuation rate estimated from $\gamma/v_g$, where $v_g$ is the group velocity of $\phi_P^{+}$ state at $k_{x}^{\bold{K}}$.}
    \label{fig.4.}
\end{figure}

To confirm these predictions, we numerically calculated the dispersion relations of the CKSs along the $x$-direction of a 1D BZ (defined as $k_x\in(-\pi/L,\pi/L)$) aligned with the domain wall separating the two domains, each consisting of $20$ unit cells. This is effectively a projection of the original 2D BZ (Fig.~\ref{fig.1.}(c)) into $k_x$ axis, see Fig.~\ref{fig.3.}(b). The two valleys of the 2D BZ correspond to $k_{x}^{\bold{K, K'}}=\pm2\pi/3L$ under this projection. As expected, the CKSs cross the bandgap near $k_{x}^{\bold{K}}$ and $k_{x}^{\bold{K'}}$ as shown in Fig.~\ref{fig.3.}(c). The phase and group velocities of the CKS supported by the P(N)-type domain wall are in the same(opposite) direction(s) as if the waves are propagating in a medium with a positive(negative) refractive index. Figure.~\ref{fig.3.}(d) depicts the confinement of CKSs at two types of domain walls.

Next, we demonstrate that valley-selective CKSs are immune to back-scattering along sharply-curved pathways. A phased array of point dipoles emitting at the mid-gap frequency $\omega_{\text{mid}}$~\cite{Supp} was used to excite a CKS in the $\bold{K}-$valley, and the non-radiative losses were modeled by assuming a finite electron scattering rate $\gamma = \omega_{\text{mid}}/300$. The robust topologically-protected propagation is observed in Figs.~\ref{fig.4.}(a)-(c). The key condition for topological protection is the valley conservation~\cite{SV2017,Photon2017}, which can be analytically proven for the $120^{\circ}$ turns (Fig.~\ref{fig.4.}(b)) using the $C^3$-symmetry of the system. On the other hand, valley conservation for the $60^{\circ}$ turn (Fig.~\ref{fig.4.}(c)) involves a more complicated mechanism because the domain walls changes between P- and N-type after the turn. At the $60^{\circ}$ turn, $\phi_{P}^{+}$ state (at $\bold{K}$) is transferred into $\phi_{N}^{+}$ state, which belongs to the $\bold{K'}$- valley with respect to the coordinate frame rotated by $60^{\circ}$, and, thus, effectively remains in the $\bold{K}$-valley in the original coordinates. Figure.~\ref{fig.4.}(d) confirms that the Drude loss is the only source of attenuation of the CKSs with or without a structural defect. Abrupt jumps in the attenuation curves at the turning points are numerical artifacts due to abrupt rotation of the integration box. Attenuation in the presence of structural defects appears to be even less than that in linear propagation, which results from overestimated propagation lengths since the CKSs don't exactly follow the prescribed zigzag domain wall~\cite{kueifu_ncomm_16}.

\begin{figure}[t]
    \includegraphics[width=0.92\columnwidth]{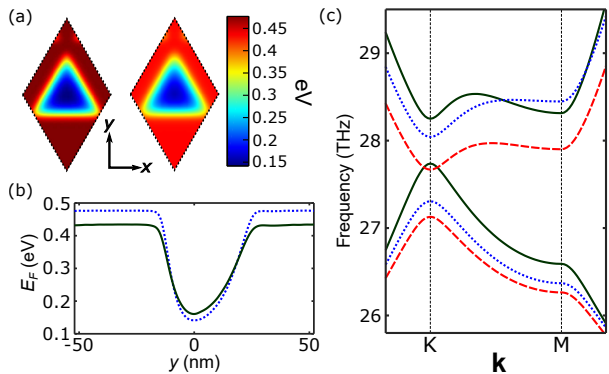}
    \caption{Non-scalable effects in hBN-encapsulated graphene ($L = \SI{60}{\nano\meter}$, $r = \SI{15}{\nano\meter}$, $h = \SI{3}{\nano\meter}$, $V=\SI{2.5}{\volt}$, and $\SI{3}{\nano\meter}$ of additional hBN layer above graphene); the hBN permittivity data is taken from Refs.~\cite{hBN1, hBN_permittivity} (a) Left: the induced Fermi energy calculated ignoring GQC, Right: GQC taken into calculation. (b) The calculated Fermi energy along $y-$axis within a unit cell calculated with(solid, dark green) and without(dotted, blue) taking GQC into consideration. (c) Bulk dispersions computed ignoring both GQC and NLER (Blue, dotted), taking only GQC into account (Red, dashed), or taking both GQC and NLER into account (Dark green, solid).}
    \label{fig.5.}
\end{figure}

Finally, we show that the non-scalable effects from GQC and NLER become significant when a realistic heterostructure---graphene encapsulated between thin hBN layers---is considered. Accounting for NLER beyond the Drude approximation requires retaining the terms at least upto the fourth-order in $\rho_2$ in the Taylor expansion of $\tilde{\chi}_{nn}\left(\bold{q},\bold{q'},\omega\right)$. For inhomogeneously doped graphene, the results obtained for the homogeneous case~\cite{RPA} were extended using the following substitutions\cite{chinn, hBN4}: $|\bold{q}|^2\rightarrow\bold{q}\cdot\bold{q'}$ and $E_F\delta_{\bold{q},\bold{q'}} \rightarrow \tilde{E}_F(\bold{q}-\bold{q'})$ (see Supplemental Material \cite{Supp} for more details). We have used the specific parameters (see caption to Fig.~\ref{fig.5.}) of the metagate in order to locate the mid-gap frequency close to one of the principal wavelength ($\lambda = \SI{10.6}{\micro\meter}$) of CO$_{2}$ lasers. First, we observe from Fig.~\ref{fig.5.}(a,b) that the depth and the sharpness of the chemical potential variation is reduced by the GQC effect. Second, according to Fig.~\ref{fig.5.}(c), the inclusion of the GQC effect red-shifts the frequencies of the GSPs. This effect is attributed to the overall reduction of $E_F$ due to GQC-related charge screening, and the dependence of $\omega_{\rm susp}$ on $E_F$. Finally, the inclusion of the NLER results in a large(comparable with the total bandgap size) blue-shift of the topological GSPs, which can be interpreted as enhanced response due to velocity matching ($\rho_2 \rightarrow 1$) between GSPs and electrons~\cite{koppens_science_17}.

In conclusion, we have described a GSP-based VC based on a deeply sub-wavelength periodic landscape of chemical potential produced by a metagate biased with respect to graphene. Highly suppressed inter-valley scattering due to extremely sub-wavelength nature of GSPs enables robust topological valley transports---guiding plasmonic CKSs along sharp curves under valley conservation. Our scheme requires neither magnetic fields nor physical patterning of graphene, and enables rapid reconfiguration by controlling the electric bias. Inherently quantum effects in graphene, such as quantum capacitance and non-local electron response, are shown to play an important role in realistic designs involving hBN encapsulation of graphene.
\begin{acknowledgments}
This work was supported by the Army Research Office (ARO) under a Grant No. W911NF-16-1-0319, and by the National Science Foundation (NSF) under the Grants No. DMR-1741788 and DMR-1719875. M.J. was also supported in part by Cornell Fellowship and in part by the Kwanjeong Fellowship from Kwanjeong Educational Foundation.
\end{acknowledgments}

\bibliography{mybib}

\clearpage

\widetext
\begin{center}
\textbf{\large Supplemental Material: Active Valley-topological Plasmonic Crystal in Metagate-tuned Graphene}\\
\end{center}
\begin{center}
Minwoo Jung$^{1}$, Zhiyuan Fan$^{2}$, and Gennady Shvets$^{2}$

	\textsl{$^{1}$Department of Physics, Cornell University, Ithaca, New York, 14853, USA}

	\textsl{$^{2}$School of Applied and Engineering Physics, Cornell University, Ithaca, New York 14853, USA}
\end{center}

\setcounter{equation}{0}
\setcounter{figure}{0}
\setcounter{table}{0}
\setcounter{page}{1}
\makeatletter
\renewcommand{\theequation}{S\arabic{equation}}
\renewcommand{\thefigure}{S\arabic{figure}}

\section{1. Numerical Method}
We elaborate on the numerical method used in this Letter and implemented in an in-house code. As briefly described in the main text (Eqs.~(1)), the method is essentially based on the linear response formulation \cite{Book1} of plasmonic excitations. For a 2D Bloch state with Bloch wavevector $\bold{k}$, the momentum integral becomes a summation over a countable set, $\bold{k} + \{\bold{G}\}$ ($\{\bold{G}\}$ is the set of the reciprocal lattice vector). Then, Eqs.~(1) in the main text are then casted into an eigenvalue problem on discrete matrices:
\begin{subequations}
\label{eq2}
\begin{gather}
\hat{V}_{\bold{k}}\hat{X}_{\bold{k}}\Ket{\phi_{n\bold{k}}} = \lambda_{n\bold{k}} \Ket{\phi_{n\bold{k}}}, \text{ with} \label{eq2a} \\
[\hat{V}_{\bold{k}}]_{\alpha, \beta} = v_{\bold{k}}\left(\bold{G}_{\alpha}, \bold{G}_{\beta},\omega\right), \text{ and} \label{eq2b}\\
[\hat{X}_{\bold{k}}]_{\alpha, \beta} = \omega^{2} \tilde{\chi}_{nn}\left(\bold{k}+\bold{G}_{\alpha},\bold{k}+\bold{G}_{\beta},\omega\right),
\end{gather}
\label{eq2}
\end{subequations}
where $\alpha$, $\beta$ are countable indices, $\bold{G}_{\alpha}, \bold{G}_{\beta} \in \{\bold{G}\}$, $\lambda_{n\bold{k}} = \omega_{n\bold{k}}^2$, and  $[\Ket{\phi_{n\bold{k}}}]_{\alpha} = \phi(\bold{k}+\bold{G}_\alpha,\omega_{n\bold{k}})$. $v_{\bold{k}}\left(\bold{q},\bold{q'},\omega\right)$ is the lattice-summed effective Coulomb interaction. We outline the properties of the Coulomb operator $\hat{V}_{\bold{k}}$ and the proper density-density response operator $\hat{X}_{\bold{k}}$, which are introduced in Eqs.~(\ref{eq2}), and discuss several computational details.

\subsection{1.1. The Coulomb operator $\hat{V}_{\bold{k}}$}
As in Eq.~(\ref{eq2b}), we define the matrix elements of the Coulomb operator to be $[\hat{V}_{\bold{k}}]_{\alpha, \beta} = v_{\bold{k}}\left(\bold{G}_{\alpha},\bold{G}_{\beta},\omega\right)$, where $v_\bold{k}\left(\bold{q},\bold{q'},\omega\right)$ is the momentum-space representation of the lattice-summed effective Coulomb interaction, $\bold{G}_{\alpha},\bold{G}_{\beta}\in\{\bold{G}\}$ ($\bold{G}$ is the reciprocal lattice vectors), and $\alpha$, $\beta$ are countable indices for $\{\bold{G}\}$. The momentum-space representation of a linear response function, $f\left(\bold{q},\bold{q'},\omega\right)$, is related to its real-space representation, $f\left(\bold{r},\bold{r'},\omega\right)$, by double Fourier transformation:

\begin{equation}
f\left(\bold{q},\bold{q'},\omega\right) = \frac{1}{|\Omega|} \int_{\Omega} d\bold{r} \int_{\Omega} d\bold{r'} e^{-i\bold{q}\cdot\bold{r}} e^{i\bold{q'}\cdot\bold{r'}}f\left(\bold{r},\bold{r'},\omega\right)
\label{eqs1}
\end{equation}
where $\Omega$ denotes the unit cell, and $|\Omega|$ is the unit cell area (if there is no periodicity in the system, $\Omega$ extends to the entire space $\mathbb{R}^{d}$, and the normalizing constant is replaced into $(2\pi)^{-d}$ with the system dimension $d$.). The real space representation of the 2D lattice-summed effective Coulomb interaction is defined as:
\begin{equation}
v_\bold{k}\left(\bold{r},\bold{r'},\omega\right) = \sum_{\bold{R}} e^{-i\bold{k}\cdot(\bold{r}-\bold{r'}-\bold{R})}v\left(\bold{r},\bold{r'}+\bold{R},\omega\right)
\label{eqs2}
\end{equation}
where $\bold{R}=n_{1}\bold{R}_1+n_{2}\bold{R}_2$ denotes the direct lattice vectors ($n_{1},n_{2}\in \mathbb{Z}$), and $v\left(\bold{r},\bold{r'},\omega\right)$ is the effective Coulomb interaction. The physical meaning of $v\left(\bold{r},\bold{r'},\omega\right)$ is the screened coulomb potential energy at position $\bold{r}$ generated by a source charge placed at position $\bold{r'}$. In this context, the screening is from surrounding dielectrics and metals, not from other electrons in the plasma. In vacuum, we simply have $v\left(\bold{r},\bold{r'},\omega\right)=e^2/4\pi \epsilon_{0} |\bold{r}-\bold{r'}|$. Thus, the Coulomb operator in vacuum is calculated to be:
\begin{equation}
\begin{aligned}
\left[\hat{V}_{\bold{k}}^{\text{vac}}\right]_{\alpha, \beta}
& =v_{\bold{k}}^{\text{vac}}\left(\bold{G}_{\alpha},\bold{G}_{\beta},\omega\right) = \frac{1}{|\Omega|} \int_{\Omega} d\bold{r} \int_{\Omega} d\bold{r'} e^{-i\bold{G}_{\alpha}\cdot\bold{r}} e^{i\bold{G}_{\beta}\cdot\bold{r'}}\sum_{\bold{R}} e^{-i\bold{k}\cdot(\bold{r}-\bold{r'}-\bold{R})}\frac{e^2}{4\pi \epsilon_{0} |\bold{r}-\bold{r'}-\bold{R}|} \\
&= \frac{1}{|\Omega|^2} \int_{\Omega} d\bold{r} \int_{\Omega} d\bold{r'} e^{-i\bold{G}_{\alpha}\cdot\bold{r}} e^{i\bold{G}_{\beta}\cdot\bold{r'}}\sum_{\bold{G}} e^{-i\bold{G}\cdot(\bold{r}-\bold{r'})}\frac{e^2}{2 \epsilon_{0} |\bold{k}-\bold{G}|}=\frac{e^2}{2 \epsilon_{0} |\bold{k}+\bold{G}_{\alpha}|}\delta_{\alpha\beta}.
\end{aligned}
\label{eqs3}
\end{equation}
We used $\sum_{\bold{R}} e^{-i\bold{k}\cdot(\bold{r}-\bold{R})}/|\bold{r}-\bold{R}|=\frac{2\pi}{|\Omega|}\sum_{\bold{G}} e^{-i\bold{G}\cdot\bold{r}}/|\bold{k}-\bold{G}|$ \cite{Ewald}. If graphene is in a homogeneous dielectric, the Coulomb operator becomes dependent of frequency due to the dispersion $\epsilon_0\rightarrow\epsilon(\omega)$. When the surrounding mediums are layered and inhomogeneous as in our system design, Eq.~(\ref{eqs1}) and Eq.~(\ref{eqs2}) are not very useful for calculating $\hat{V}_{\bold{k}}$. Instead, we start from a Bloch ansatz of the plasma field, $\Phi_{\bold{k}}(\bold{r}, z)$ ($\bold{r}=x\hat{\bold{x}}+y\hat{\bold{y}}$), in an electrostatic limit ($\nabla^2\Phi_{\bold{k}}=0$):
\begin{equation}
\Phi_{\bold{k}}(\bold{r},z) = \begin{cases}
\sum_{\alpha} e^{i\bold{k}_{\alpha}\cdot\bold{r}}\left(c_{\alpha}\cosh(|\bold{k}_{\alpha}|z) -s_{\alpha} \sinh(|\bold{k}_{\alpha}|z)\right) & (z\geq0)\\
\sum_{\alpha} e^{i\bold{k}_{\alpha}\cdot\bold{r}}\left(c_{\alpha}\cosh\left(|\bold{k}_{\alpha}|z\right) +s'_{\alpha} \sinh\left(|\bold{k}_{\alpha}|z\right)\right) & (-h\leq z\leq0)
\end{cases}
\text{    }(\bold{k}_\alpha \equiv \bold{k}+\bold{G}_\alpha);
\label{eqs4}
\end{equation}
here, graphene is placed at $z=0$, and MG is placed at $z=-h$. Then, following the vector representation of state vectors introduced in Eqs.~(2), the vector elements of total (screened) potential on graphene $\ket{\phi_{\bold{k}}}$ and induced charge density on graphene $\ket{n_{\bold{k}}}$ are given as $[\ket{\phi_{\bold{k}}}]_\alpha = c_{\alpha}$ and $[\ket{n_{\bold{k}}}]_\alpha = \frac{\epsilon_0}{e^2}|\bold{k}_{\alpha}|(s_{\alpha}+s'_{\alpha})$. If we find relations between $c_\alpha$ and $s_\alpha, s'_\alpha$ by matching boundary conditions, we can derive the expression for $\hat{V}_{\bold{k}}$ from $\ket{\phi_{\bold{k}}}=\hat{V}_{\bold{k}}\ket{n_{\bold{k}}}$.

For this section, we assume that MG is infinitely deep (a few times $h$ suffices this assumption) and the permittivity of the metal is infinite. For an unstructured metal gate, we have simple boundary conditions: $\Phi_{\bold{k}}(\bold{r},z=\infty)=\Phi_{\bold{k}}(\bold{r},z=-h)=0$. Then, we get $c_\alpha=s_\alpha=\tanh(|\bold{k}_{\alpha}|h)s'_\alpha$ and, thus, $[\hat{V}_{\bold{k}}^{\text{flat}}]_{\alpha, \beta}=\frac{e^2\delta_{\alpha\beta}}{\epsilon_{0} |\bold{k}_{\alpha}|(1+\coth(|\bold{k}_{\alpha}|h))}$. As pointed out in \cite{hBN3}, the effective coulomb interaction does not diverge even at long-wavelength limit due to the screening from the metal gate: $\lim_{\bold{k}_\alpha\rightarrow0}[\hat{V}_{\bold{k}}]_{\alpha, \alpha}=e^2h/\epsilon_0$. Now, we turn into our design where the metal gate is periodically carved out with triangular cylindrical holes. We use the set of laplacian eigenmodes in a triangular cylinder, $T_\mu(\bold{r})$, \cite{Book2} to describe the field in the MG domain ($z<-h$):
\begin{equation}
\begin{split}
\Phi_{\bold{k}}(\bold{r},z<-h) =\sum_{\bold{R}} e^{i\bold{k}\cdot\bold{R}}\phi_{\text{hole}}\left(\bold{r}-\bold{R},z\right) \text{, where }
\phi_{\text{hole}}\left(\bold{r},z\right) = \sum_{\mu}t_{\mu}T_{\mu}\left(\bold{r}\right)e^{\xi_{\mu}(z+h)},
\end{split}
\label{eqs5}
\end{equation}
and $\mu$ is a countable index for the laplacian eigenmodes. $\xi_\mu$ is given to satisfy $\nabla^{2}_{\text{2D}}T_\mu(\bold{r})+\xi_\mu^2T_\mu(\bold{r})=0$. $T_\mu(\bold{r})$ vanishes outside of the hole, and forms an eigenbasis inside the hole: $\int_S T_{\mu}^{*}(\bold{r}) T_{\nu}(\bold{r}) d\bold{r}=|S|\delta_{\mu\nu}$ ($S$ denotes the hole region, and $|S|$ is its section area). For compact description of boundary conditions, we use the vector representation for the linear coefficients: $\ket{c}\equiv\{c_\alpha\}, \ket{s'}\equiv\{s'_\alpha\}$, and $\ket{t}\equiv\{t_\mu\}$. Also, we take several groups of repeatedly used numbers in operator forms: $[\hat{C}_\bold{k}]_{\alpha, \beta}=\cosh(|\bold{k_\alpha}|h)\delta_{\alpha\beta}$, $[\hat{S}_\bold{k}]_{\alpha, \beta}=\sinh(|\bold{k_\alpha}|h)\delta_{\alpha\beta}$, $[\hat{K}_\bold{k}]_{\alpha, \beta}=|\bold{k_\alpha}|\delta_{\alpha\beta}$, $[\hat{P}]_{\alpha, \beta}=\frac{1}{|\Omega|}\int_S e^{-i(\bold{k}_\alpha-\bold{k}_\beta)\cdot\bold{r}}d\bold{r}$, $[\hat{B}_\bold{k}]_{\mu, \alpha}=\frac{1}{|S|}\int_S T_{\mu}^{*}(\bold{r}) e^{i\bold{k}_\alpha\cdot\bold{r}}d\bold{r}$, and $[\hat{\Xi}]_{\mu, \nu}=\xi_\mu \delta_{\mu\nu}$. $\hat{P}$ does a spatial projection into the hole region $S$, and $\hat{B}_\bold{k}$ is the basis transformation from $\{e^{i\bold{k}_\alpha \cdot \bold{r}}\}$ into $\{T_\mu(\bold{r})\}$. Then, the boundary conditions at $z=-h$ are expressed as:
\begin{equation}
\begin{split}
\hat{P}\left(\hat{C}_{\bold{k}}\ket{c}-\hat{S}_{\bold{k}}\ket{s'}\right)=\hat{C}_{\bold{k}}\ket{c}-\hat{S}_{\bold{k}}\ket{s'} \text{ }\text{ }\text{ }\text{ }&(\because \Phi_{\bold{k}}(\bold{r},z=-h) \text{ vanishes outside the hole.}),\\
\ket{t} = \hat{B}_{\bold{k}}\left(\hat{C}_{\bold{k}}\ket{c}-\hat{S}_{\bold{k}}\ket{s'}\right) \text{ }\text{ }\text{ }\text{ }&(\because\text{Continuity of }\Phi_{\bold{k}}(\bold{r},z) \text{ at }z=-h),\\
\hat{\Xi}\ket{t} = \hat{B}_{\bold{k}}\left(\hat{C}_{\bold{k}}\hat{K}_{\bold{k}}\ket{s'}-\hat{S}_{\bold{k}}\hat{K}_{\bold{k}}\ket{c}\right) \text{ }\text{ }\text{ }\text{ }&(\because\text{Continuity of }\partial_z\Phi_{\bold{k}}(\bold{r},z) \text{ at }z=-h).
\end{split}
\label{eqs6}
\end{equation}
Using $\frac{1}{|S|}\sum_{\mu}T_{\mu}^{*}\left(\bold{r'}\right)T_{\mu}\left(\bold{r}\right)=\delta(\bold{r}-\bold{r'})$, we get $\hat{B}_{\bold{k}}^{\dagger}\hat{B}_{\bold{k}}=\frac{|\Omega|}{|S|}\hat{P}$. Then, Eqs.~(\ref{eqs6}) is simplified to:
\begin{equation}
\begin{split}
\hat{C}_{\bold{k}}\ket{c}-\hat{S}_{\bold{k}}\ket{s'}&=\hat{Q}_{\bold{k}}\left(\hat{C}_{\bold{k}}\hat{K}_{\bold{k}}\ket{s'}-\hat{S}_{\bold{k}}\hat{K}_{\bold{k}}\ket{c}\right)
\longrightarrow \ket{s'}=\left(\hat{Q}_{\bold{k}}\hat{C}_{\bold{k}}\hat{K}_{\bold{k}}+\hat{S}_{\bold{k}}\right)^{-1}\left(\hat{Q}_{\bold{k}}\hat{S}_{\bold{k}}\hat{K}_{\bold{k}}+\hat{C}_{\bold{k}}\right)\ket{c}\\
&\xrightarrow{\hat{C}_{\bold{k}}^2-\hat{S}_{\bold{k}}^2=\bold{I}} \hat{K}_{\bold{k}}\ket{s'}=\left(\hat{K}_{\bold{k}}\hat{S}_{\bold{k}}\hat{C}_{\bold{k}}^{-1}+\left(\hat{C}_{\bold{k}}\hat{Q}_{\bold{k}}\hat{C}_{\bold{k}} + \hat{K}_{\bold{k}}^{-1}\hat{S}_{\bold{k}}\hat{C}_{\bold{k}}\right)^{-1}\right)\ket{c},
\end{split}
\label{eqs7}
\end{equation}
where $\hat{Q}_{\bold{k}}\equiv\frac{|S|}{|\Omega|}\hat{B}_{\bold{k}}^{\dagger}\hat{\Xi}^{-1}\hat{B}_{\bold{k}}$. The matrix $\hat{Q}_{\bold{k}}$ can be directly calculated as cross-Fourier transform of the green's function inside the holes $g(\bold{r},\bold{r'})=\frac{1}{|S|}\sum_\mu \xi_\mu^{-1}T_{\mu}(\bold{r}) T_{\mu}^{*}(\bold{r'})$ \cite{Book2}:
\begin{equation}
\begin{split}
\left[\hat{Q}_{\bold{k}}\right]_{\alpha, \beta} = \left[\frac{|S|}{|\Omega|}\hat{B}_{\bold{k}}^{\dagger}\hat{\Xi}^{-1}\hat{B}_{\bold{k}}\right]_{\alpha, \beta} &= \frac{|S|}{|\Omega|}\sum_\mu \xi_\mu^{-1} \frac{1}{|S|}\int_S T_{\mu}(\bold{r}) e^{-i\bold{k}_\alpha\cdot\bold{r}}d\bold{r}  \frac{1}{|S|}\int_S T_{\mu}^{*}(\bold{r'}) e^{i\bold{k}_\beta\cdot\bold{r'}}d\bold{r'}\\
&=\frac{1}{|\Omega|}\int_S d\bold{r}\int_S d\bold{r'} e^{-i\bold{k}_\alpha\cdot\bold{r}}e^{i\bold{k}_\beta\cdot\bold{r'}} g(\bold{r},\bold{r'}).
\label{eqs10}
\end{split}
\end{equation}
Recalling $\ket{c}=\frac{\epsilon_0}{e^2}\hat{V}_{\bold{k}}\hat{K}_{\bold{k}}\left(\ket{s}+\ket{s'}\right)$, we finally obtain the exact expression for $\hat{V}_{\bold{k}}$ that captures the MG screening effect:
\begin{equation}
\hat{V}_{\bold{k}}^{\text{MG}}=\frac{e^2}{\epsilon_0}\left(\hat{K}_{\bold{k}}+\hat{K}_{\bold{k}}\hat{S}_{\bold{k}}\hat{C}_{\bold{k}}^{-1}+\left(\hat{C}_{\bold{k}}\hat{Q}_{\bold{k}}\hat{C}_{\bold{k}} + \hat{K}_{\bold{k}}^{-1}\hat{S}_{\bold{k}}\hat{C}_{\bold{k}}\right)^{-1}\right)^{-1} .
\label{eqs8}
\end{equation}
As expected, $\hat{V}_{\bold{k}}^{\text{MG}}$ is Hermitian, because $\hat{Q}_{\bold{k}}$, $\hat{K}_{\bold{k}}$ ,$\hat{S}_{\bold{k}}$, and $\hat{C}_{\bold{k}}$ are Hermitian. When the hole is removed ($|S|\rightarrow0$ and, thus, $\hat{Q}_{\bold{k}}\rightarrow\bold{0}$), $\hat{V}_{\bold{k}}^{\text{MG}}$ goes back to $\hat{V}_{\bold{k}}^{\text{flat}}$.

\begin{figure}
    \includegraphics[width=0.8\columnwidth]{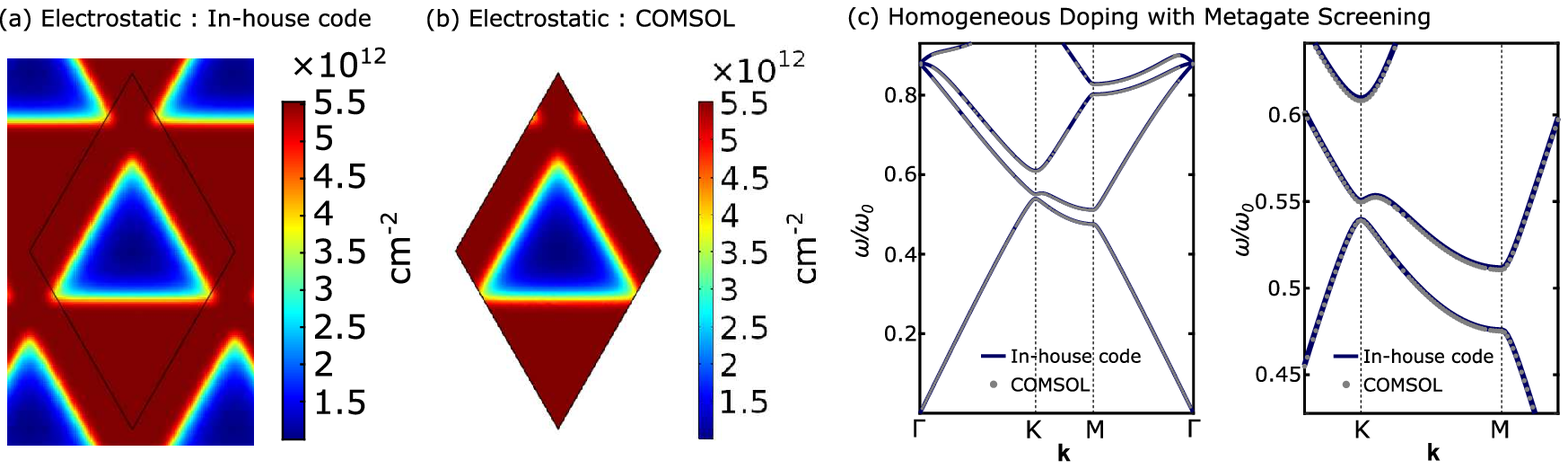}
    \caption{Verification of the validity of Eq.~(\ref{eqs8}). (a) Induced charge density computed by Eq.~(\ref{eqs9}) with $\alpha = 30^{\circ}$, $L/r = 4, L/h = 25$ and $V/h = \SI{1}{\volt/\nano\meter}$. (b) Induced charge density computed by COMSOL Electrostatic module with the same parameter conditions used in (a). (c) Left: Plasmonics dispersion calculated by our in-house code using Eq.~(\ref{eqs8}) (dark blue curves) and by COMSOL Electromagnetic Waves, Frequency Domain module (gray dots); we impose a homogeneous doping in order to highlight the effect of MG screening ($\alpha = 30^{\circ}$, $L/r = 4, L/h = 16$). Right: Zoom on the bandgaps.}
    \label{fig.S1.}
\end{figure}

We confirm that the complicated expression given in Eq.~(\ref{eqs8}) actually works well, by cross-checking the numerical results from Eq.~(\ref{eqs8}) and COMSOL Multiphysics. First, we compare the results of electrostatic simulations. When a static voltage difference, $V$, is applied between graphene and MG, Fourier coefficients of the induced charge density, $\ket{n_{ES}}$, is simply obtained by:
\begin{equation}
\ket{n_{ES}}=\left(\hat{V}_{\bold{k}\rightarrow0}^{\text{MG}}\right)^{-1}\ket{\phi_{ES}},
\label{eqs9}
\end{equation}
where $[\ket{\phi_{ES}}]_\alpha = eV\delta_{\alpha0}$ ($\bold{G}_0 = \bold{0}$). Figure.~\ref{fig.S1.}(a) illustrates the spatial profile of the induced charge density obtained by taking an inverse-Fourier transform of $\ket{n_{ES}}$, and it exactly matches with the result from COMSOL Electrostatic module (Fig.~\ref{fig.S1.}(b)). For electrostatic simulation, we take both the Fourier coefficients and the linear coefficients of the laplacian eigenmodes upto the 15$^{\text{th}}$ order ($|n_1|,|n_2|\leq15$ with $\bold{G}=n_1\bold{G}_1 + n_2\bold{G}_2$, and $0< p,q\leq15$ with $T_{\mu=(p,q)}(\bold{r})$). Second, we assume a homogeneous doping, $E_F({\bold{r}})=E_0$, and investigate the effect of MG screening in plasmonic dispersion. For comparison with COMSOL Multiphysics, we use Electromagnetic Waves, Frequency Domain module, where graphene can be imitated by exploiting surface current density. Figure.~\ref{fig.S1.}(c) depicts the calculated plasmonic dispersions; again, two simulations show a good agreement. Here, the scaling frequency $\omega_0$ is given as $\omega_{0} = \sqrt{e^2 E_{0}/\epsilon_0 \hbar^2 L}$. We observe that, due to the periodic structure in MG, bandgaps are opened at $\bold{{K}}$ and $\bold{M}$ points even under homogeneous doping. We don't use COMSOL Multiphysics for other simulations because (1) COMSOL creates a lot of spurious modes that need to be removed manually, (2) it even misses some of the physical eigenmodes as represented by irregular appearance of gray dots in Fig.~\ref{fig.S1.}(c), and (3) the computation is much slower (more than 100-fold) than our in-house codes. This drastic difference in computational times comes from the fact that our method treats the system still as 2D by capturing all inhomogeneity along $z$-axis into 2D effective Coulomb interaction, whereas COMSOL runs finite element methods in a 3D system.

Actually, Eqs.~(\ref{eqs5})$\sim$(\ref{eqs9}) don't assume a specific shape of the holes. Thus, this method can be easily applied to circular, rectangular, or hexagonal holes, in which the analytic expression for laplacian eigenmodes are well-known.

\subsection{1.2. The density-density response operator $\hat{X}_{\bold{k}}$}
We start from evaluating $\hat{X}_{\bold{k}}$ in local Drude approximation \cite{Drude}: $\sigma_{\text{Drude}}\left(\bold{r},\bold{r'},\omega\right)=$$\frac{e^2}{\pi \hbar^2} \frac{i}{\omega+i\gamma}$$E_{F}\left(\bold{r}\right)\delta(\bold{r}-\bold{r'})$; then,	 Eq.~(\ref{eqs1}) gives

\begin{equation}
\sigma_{\text{Drude}}\left(\bold{q},\bold{q'},\omega\right) = \frac{1}{|\Omega|} \int_{\Omega} d\bold{r} \int_{\Omega} d\bold{r'} e^{-i\bold{q}\cdot\bold{r}} e^{i\bold{q'}\cdot\bold{r'}}\frac{e^2}{\pi \hbar^2} \frac{i}{\omega+i\gamma}E_{F}\left(\bold{r}\right)\delta(\bold{r}-\bold{r'})=\frac{e^2}{\pi \hbar^2} \frac{i}{\omega+i\gamma}\tilde{E}_{F}\left(\bold{q}-\bold{q'}\right),
\label{eqs11}
\end{equation}
where $\tilde{E}_{F}(\bold{q})=\frac{1}{|\Omega|}\int_\Omega e^{-i\bold{q}\cdot\bold{r}}E_{F}(\bold{r})d\bold{r}$. Recalling the definition of the proper density-density response function $n\left(\bold{q},\omega\right) = \sum_{\bold{q'}}\tilde{\chi}_{nn}\left(\bold{q},\bold{q'},\omega\right)\phi\left(\bold{q'},\omega\right)$ and using a set of relations, (1) $\bold{J}\left(\bold{q},\omega\right)=\sum_{\bold{q'}}\sigma\left(\bold{q},\bold{q'},\omega\right)\bold{E}\left(\bold{q'},\omega\right)$, (2) $i\bold{q}\cdot\bold{J}\left(\bold{q},\omega\right)-i\omega (-e)n\left(\bold{q},\omega\right)= 0$, and (3) $-e\bold{E}\left(\bold{q},\omega\right)=-i\bold{q}\phi\left(\bold{q},\omega\right)$, we obtain the fundamental relation between the conductivity and the proper density-density response function: $\tilde{\chi}_{nn}\left(\bold{q},\bold{q'},\omega\right) = \frac{\bold{q}\cdot\bold{q'}}{i\omega e^2}\sigma\left(\bold{q},\bold{q'},\omega\right)$. Thus, $\hat{X}_{\bold{k}}$ in local Drude approximation is given as
\begin{equation}
[\hat{X}_{\bold{k}}^{\text{Drude}}]_{\alpha, \beta} = \omega(\omega+i\gamma) \tilde{\chi}_{nn}^{\text{Drude}}\left(\bold{k}_\alpha,\bold{k}_\beta,\omega\right) = \frac{\bold{k}_{\alpha}\cdot\bold{k}_{\beta}\tilde{E}_{F}\left(\bold{k}_{\alpha}-\bold{k}_{\beta}\right)}{\pi\hbar^2}.
\label{eqs12}
\end{equation}
As mentioned in the main text, $\omega^2$ is substituted by $\omega(\omega+i\gamma)$ as the Drude loss $\gamma$ is included in the analysis.

A more precise description of $\tilde{\chi}_{nn}\left(\bold{q},\bold{q'},\omega\right)$ beyond Drude approximation is obtained in Random Phase Approximation (RPA), in which the proper density-density response function is simply replaced by the Hartree response function: $\tilde{\chi}_{nn}^{\text{RPA}}\left(\bold{q},\bold{q'},\omega\right)=\chi_{\text{H}}\left(\bold{q},\bold{q'},\omega\right)$ \cite{Book1}. In a homogeneous electron gas, the Hartree response function reduces to the density-density response function in non-interacting limit, and the density-density response function of non-interacting 2D massless electron gas has an exact analytic expression in a non-relativistic limit \cite{RPA}:
\begin{equation}
\begin{split}
\chi_{\text{H}}^{\text{hom}}\left(\bold{q},\bold{q'},\omega\right) \cong \delta_{\bold{q},\bold{q'}}\frac{E_F}{\pi \hbar^2 v_F^2}\left(\frac{x^2y}{\sqrt{1-x^2}}\left(F_{+}(x,y)-F_{-}(x,y)\right) - 2\right),\text{ where} \\
x\equiv v_F |\bold{q}|/\omega,\text{ }, y\equiv \hbar \omega / 2E_F\text{, and }F_{\pm}(x,y)=\int_1^{\frac{1\pm y}{xy}} du \sqrt{1-u^2};
\end{split}
\end{equation}
here, we assume $x<1$ (suppressed intraband transitions), $y(1+x)<1$ (suppressed interband transitions), and the zero temperature limit ($2E_F \gg k_B T$). For our purpose, we introduce the taylor expansion of $\chi_{\text{H}}^{\text{hom}}\left(\bold{q},\bold{q'},\omega\right)$ of a homogeneous 2D massless electrons respect to $x$ and $y$ upto the 6$^{\text{th}}$ order terms: 
\begin{equation}
\begin{split}
\chi_{\text{H}}^{\text{hom}}\left(\bold{q},\bold{q'},\omega\right) &= \delta_{\bold{q},\bold{q'}}\frac{E_F}{\pi \hbar^2 v_F^2}\left(x^2 - x^2y^2 + \frac{3}{4}x^4 -\frac{1}{3}x^2y^4 - \frac{1}{2}x^4y^2 + \frac{5}{8}x^6 + O(x^n y^m; n+m>6)\right) \\
&=\delta_{\bold{q},\bold{q'}}\left(\frac{E_{F}|\bold{q}|^2}{\pi\hbar^2\omega^2}-\frac{ E_F^{-1}|\bold{q}|^2}{4\pi}+ \frac{3v_{F}^{2}E_{F}|\bold{q}|^4}{4\pi\hbar^2\omega^4}-\frac{\hbar^2 \omega^2  E_F^{-3}|\bold{q}|^2}{48\pi} -\frac{v_{F}^{2} E_F^{-1}|\bold{q}|^4}{8\pi\omega^2}+ \frac{5v_{F}^{4}E_{F}|\bold{q}|^6}{8\pi\hbar^2\omega^6} + O(x^n y^m; n+m>6)\right).
\end{split}
\label{eqs13}
\end{equation}
The Hartree response function in inhomogeneously doped graphene, however, has not been microscopically calculated due to the difficulty in quantifying the interband transitions with a spatially varying Fermi level \cite{chinn}. Instead, Ref.~\cite{chinn} suggested a semi-phenomenological expression for $\chi_{\text{H}}^{4th}\left(\bold{q},\bold{q'},\omega\right)$ in an inhomogeneous case at small $x$ and $y$:
\begin{equation}
\chi_{\text{H}}^{4th}\left(\bold{q},\bold{q'},\omega\right) = \tilde{\chi}_{nn}^{\text{Drude}}\left(\bold{q},\bold{q'},\omega\right) -\frac{\tilde{E}^{-1}_{F}\left(\bold{q}-\bold{q'}\right)\bold{q}\cdot\bold{q'}}{4\pi}+ \frac{3v_{F}^{2}\tilde{E}_{F}\left(\bold{q}-\bold{q'}\right)\left(\bold{q}\cdot\bold{q'}\right)^2}{4\pi\hbar^2\omega^4},
\label{eqs14}
\end{equation}
where $\tilde{E}_{F}^{n}(\bold{q})=\frac{1}{|\Omega|}\int_\Omega e^{-i\bold{q}\cdot\bold{r}}E_{F}^{n}(\bold{r})d\bold{r}$. This expression is acquired by substituting $\delta_{\bold{q},\bold{q'}}E_F$ to $\tilde{E}_{F}\left(\bold{q}-\bold{q'}\right)$, $\delta_{\bold{q},\bold{q'}}E_F^{-1}$ to $\tilde{E}_{F}^{-1}\left(\bold{q}-\bold{q'}\right)$, and $|\bold{q}|^2$ to $\bold{q}\cdot\bold{q'}$ from Eq.~(\ref{eqs13}), so that it reduces back to Eq.~(\ref{eqs13}) in homogeneous limit while minimally capturing the non-local effects through the fourth-order dependence on momentum in the third term. Eq.~(\ref{eqs14}) was successfully used in \cite{hBN4} to predict the plasmonic scattering in an inhomogeneously doped graphene. We also use Eq.~(\ref{eqs14}) to examine the significance of non-local effects plasmonic dispersions in Fig. 5 in the main text and in the section $\bold{2.1.}$ of this Supplemental Material. Similarly, we can extend this semi-phenomenological expression upto the 6th-order terms:
\begin{equation}
\chi_{\text{H}}^{6th}\left(\bold{q},\bold{q'},\omega\right) = \chi_{\text{H}}^{4th}\left(\bold{q},\bold{q'},\omega\right) -\frac{\hbar^2 \omega^2 \tilde{E}_F^{-3}(\bold{q}-\bold{q'})\bold{q}\cdot\bold{q'}}{48\pi} -\frac{v_{F}^{2} \tilde{E}_F^{-1}(\bold{q}-\bold{q'})(\bold{q}\cdot\bold{q'})^2}{8\pi\omega^2}+ \frac{5v_{F}^{4}\tilde{E}_{F}(\bold{q}-\bold{q'})(\bold{q}\cdot\bold{q'})^3}{8\pi\hbar^2\omega^6}.
\label{eqs14_2}
\end{equation}

\begin{figure}[t]
    \includegraphics[width=0.8\columnwidth]{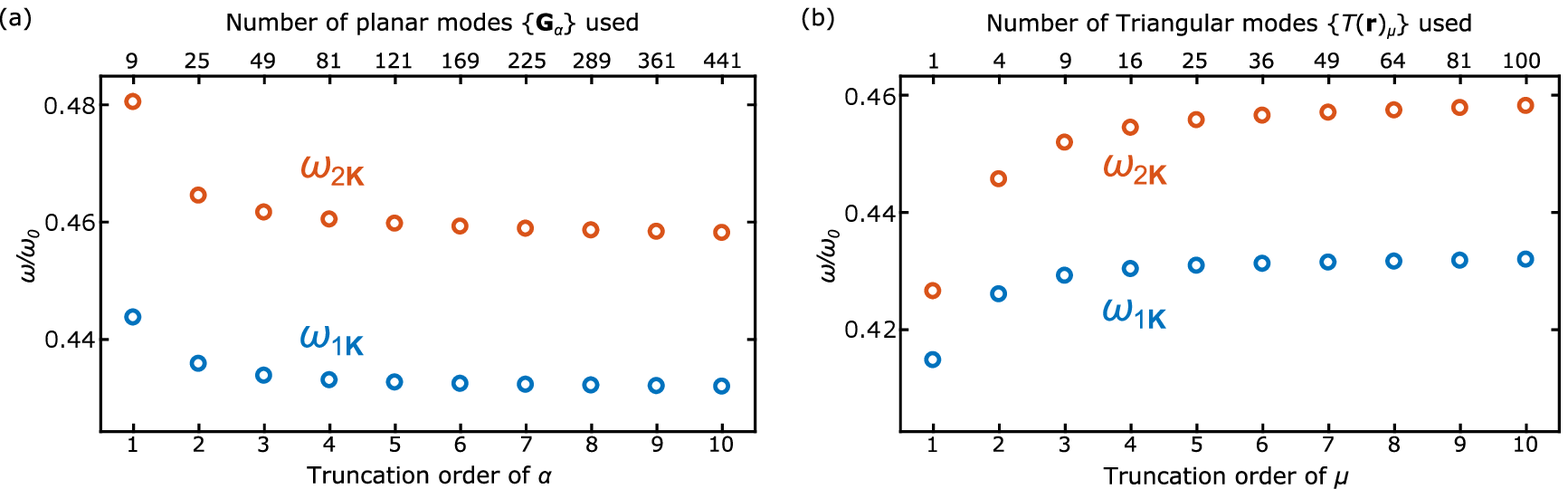}
    \caption{Convergence of the dispersion calculation vs. truncation orders ($L/r=4$, $L/h=25$, $\theta=30^\circ$). (a) The convergence of calculated frequencies at $\bold{K}$ point below and above the bandgap; the truncation order of $\mu$ is fixed at 10. (b) Contrary to (a), the truncation order of $\alpha$ is fixed at 10, and the truncation order of $\mu$ is varying.}
    \label{fig.S2.}
\end{figure}

\subsection{1.3. 2D bulk dispersion}
The eigenvalue problem given in Eq.~(\ref{eq2a}) is generally nonlinear, because the operators themselves are not independent of the eigenvalues. Only in local Drude approximation and in vacuum, as mentioned above, both $\hat{V}_{\bold{k}}$ (Eq.~(\ref{eqs8}) and $\hat{X}_{\bold{k}}$ (Eq.~(\ref{eqs12})) are independent of the frequency, and Eq.~(\ref{eq2a}) becomes an ordinary linear eigenvalue problem. $\hat{X}_{\bold{k}}$ and $\hat{V}_{\bold{k}}$ are discrete, but infinite matrices; thus, we need to truncate the matrices beyond certain orders in numerical calculations. Figure.~\ref{fig.S2.} depicts the speed of convergence of the numerical results. If the truncation order of $\alpha$ is $N$, we use $(2N+1)^2$ planar modes; $|n_1|,|n_2|\leq N$ with $\bold{G}_{\alpha = (n_1,n_2)}=n_1\bold{G}_1 + n_2\bold{G}_2$. Similarly, if the truncation order of $\mu$ is $M$, we use $M^2$ laplacian eigenmodes of a triangular hole for computing $\hat{Q}_\bold{k}$; $0< p,q\leq M$ with $T_{\mu=(p,q)}(\bold{r})$. We use $N=M=10$ in calculating the bulk dispersions of Fig.~1(d)-(e) and Fig.~2(b) in the main text.

When we use Eq.~(\ref{eqs14}) for the density-density response operator, instead of Eq.~(\ref{eqs12}), for evaluating the non-local effects in the section $\bold{2.1.}$, the eigenvalue problem obtains a quadratic form:
\begin{equation}
\left(\lambda_{n\bold{k}}^2\left(\hat{X}_{\bold{k}}^{(2)}\hat{V}_{\bold{k}}^{\text{MG}}-\bold{I}\right) + \lambda_{n\bold{k}}\hat{X}_{\bold{k}}^{(1)}\hat{V}_{\bold{k}}^{\text{MG}}+\hat{X}_{\bold{k}}^{(3)}\hat{V}_{\bold{k}}^{\text{MG}}\right)\Ket{\phi_{n\bold{k}}}=0,
\label{eqs15}
\end{equation}
where $\hat{X}_{\bold{k}}^{(1)}=\hat{X}_{\bold{k}}^{\text{Drude}}$, $[\hat{X}_{\bold{k}}^{(2)}]_{\alpha,\beta}=-\frac{1}{4\pi}\tilde{E}_{F}^{-1}\left(\bold{k}_{\alpha}-\bold{k}_{\beta}\right)\bold{k}_{\alpha}\cdot\bold{k}_{\beta}$, and $[\hat{X}_{\bold{k}}^{(3)}]_{\alpha,\beta}=\frac{3v_F^2}{4\pi\hbar^2}\tilde{E}_{F}\left(\bold{k}_{\alpha}-\bold{k}_{\beta}\right)\left(\bold{k}_{\alpha}\cdot\bold{k}_{\beta}\right)^2$. We solve this quadratic eigenvalue equation for calculating plasmonic dispersions including the non-local effects in Fig.~\ref{fig.S5.}(a).

For the calculation of Fig.~5 in the main text, $\hat{V}_\bold{k}^{\text{MG}}$ is slightly modified to account for the anisotropic($\epsilon_{||}\neq\epsilon_{\perp}$) dispersion of hBN:
\begin{equation}
\hat{V}_{\bold{k}}^{\text{MG}}=\frac{e^2}{\epsilon_{\text{eff}}}\left(\hat{K}_{\bold{k}}+\hat{K}_{\bold{k}}\hat{S'}_{\bold{k}}\hat{C'}_{\bold{k}}^{-1}+\left(\hat{C'}_{\bold{k}}\epsilon_{\text{eff}}\hat{Q}_{\bold{k}}\hat{C'}_{\bold{k}} + \hat{K}_{\bold{k}}^{-1}\hat{S'}_{\bold{k}}\hat{C'}_{\bold{k}}\right)^{-1}\right)^{-1},
\label{eqshbn}
\end{equation}
where $[\hat{S'}_{\bold{k}}]_{\alpha, \beta}=\sinh(\eta|\bold{k_\alpha}|h)\delta_{\alpha\beta}$, $[\hat{C'}_{\bold{k}}]_{\alpha, \beta}=\cosh(\eta|\bold{k_\alpha}|h)\delta_{\alpha\beta}$, $\epsilon_{\text{eff}}=\sqrt{\epsilon_{||}\epsilon_{\perp}}$, and $\eta=\sqrt{\epsilon_{||}/\epsilon_{\perp}}$. Then, $\hat{V}_{\bold{k}}^{\text{MG}}$ acquires non-polynomial frequency dependence due to the dispersive nature of the permittivity. This non-polynomial type of eigenvalue problems is then solved by iterations just like in the Newton's method.

\begin{figure}[b]
    \includegraphics[width=0.8\columnwidth]{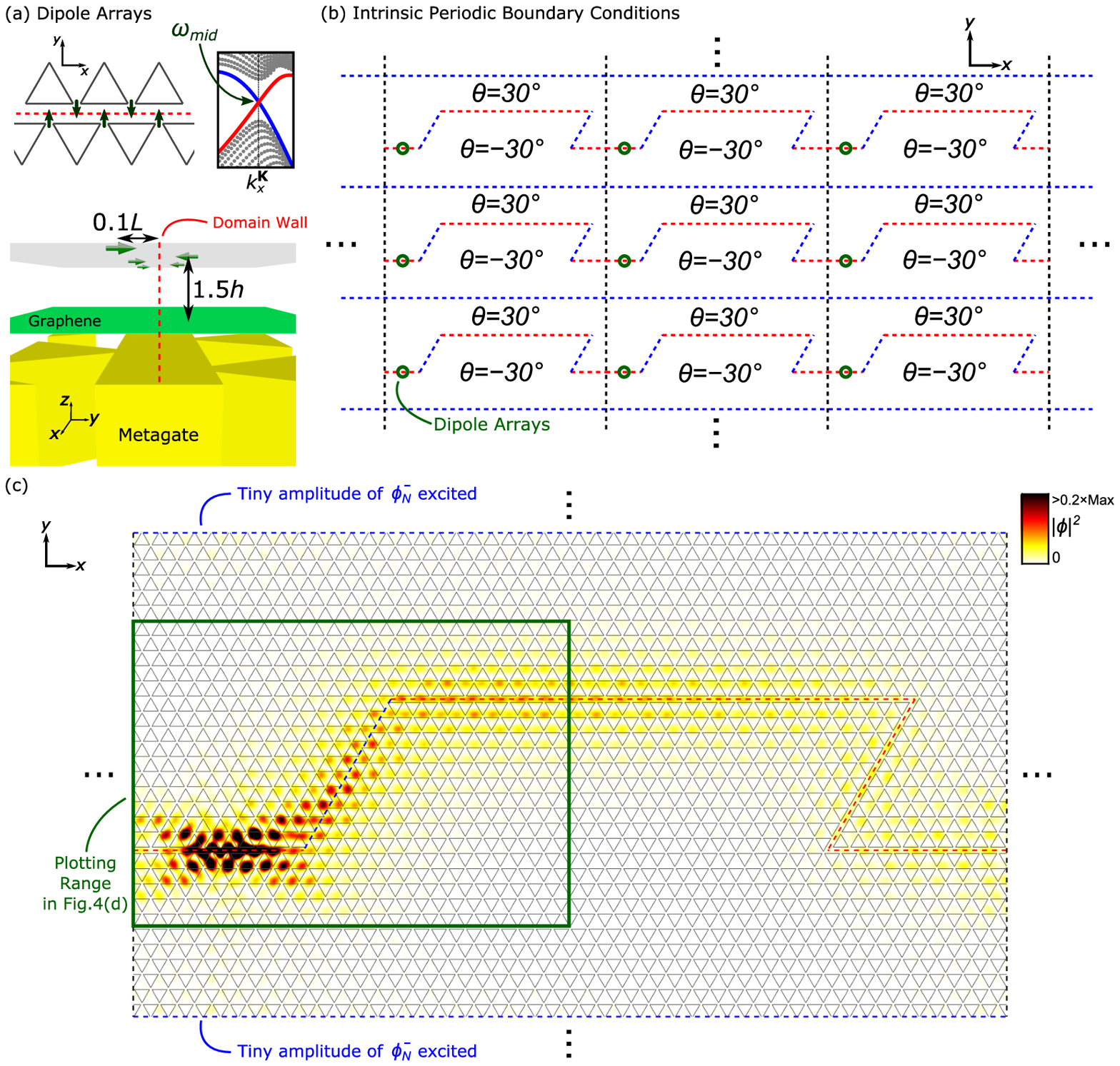}
    \caption{Computational details in the calculations of Fig.~4. (a) Dipole array. (b) Illustration of the periodic boudary conditions in the example of Fig.~4(d); the positive-type domain walls are in red dotted lines, and the negative-type domain walls are in blue dotted lines (the black dotted lines are borders of unit cells). (c) Full computational geometry of Fig.~4(d).}
    \label{fig.S3.}
\end{figure}

\subsection{1.4. 1D Edge dispersion}
In the edge dispersion calculation of Fig.~3(c), we essentially perform a 2D bulk dispersion calculation with a nanoribbon-like unit cell that consists of the upper domain of 40 up-triangles ($\theta=30^\circ$) and the lower domain of 40 down-triangles ($\theta=-30^\circ$), see Fig.~3(b). The length of the unit cell is $L_x = L$ along the $x$-axis, and $L_y = 20\sqrt{3}L$ along the $y$-axis. The $y$-component of Bloch wavevector is taken to be zero, since we are interested only in the periodicity along the $x$-axis. Within the unit cell, the linear coefficients $t_\mu$ of of each hole are completely independent. Thus, Eq.~(\ref{eqs5}) needs to be modified as:
\begin{equation}
\begin{split}
\Phi_{\bold{k}=k_x\hat{\bold{x}}}&(\bold{r},z<-h) =\sum_{\bold{R}} e^{i \bold{k}\cdot\bold{R}}\sum_{j}\phi_{\text{hole}}^j\left(\bold{r}-\bold{r}_c^j -\bold{R},z\right) \text{, with }\\
\phi_{\text{hole}}^j\left(\bold{r},z\right) & = \sum_{\mu}t_{\mu}^j T_{\mu}^j\left(\bold{r}\right)e^{\xi_{\mu}(z+h)} \text{ and } T_{\mu}^j\left(x,y\right)=\begin{cases}T_{\mu}\left(x,y\right) & (\theta_j = 30^\circ)\\T_{\mu}\left(x,-y\right) & (\theta_j = -30^\circ)\end{cases}
\end{split}
\label{eqs16}
\end{equation}
where $j$ is the index for 80 different triangular holes within a unit cell, $\bold{r}_c^j$ is the relative displacement of the center of the hole $j$ from the origin of the unit cell, and $\theta_j$ is the rotation angle of the hole $j$. Here, $T_\mu(\bold{r})$ specifically represents the triangular eigenmodes of a triangular hole with the orientation $\theta=30^\circ$. Then, the matrix elements of the basis transformation operator $\hat{B}_{\bold{k}=k_x\hat{\bold{x}}}$ changes accordingly:
\begin{equation}
[\hat{B}_{\bold{k}=k_x\hat{\bold{x}}}]_{\left(\mu,j\right) \alpha}=\frac{1}{|S|}\int_S T_{\mu}^{j*}(\bold{r}-\bold{r}_c^j) e^{i\bold{k}_\alpha\cdot\bold{r}}d\bold{r}.
\label{eqs17}
\end{equation}
Here, the number pair of $\mu$ and $j$ comprises of the row index of $\hat{B}_{\bold{k}=k_x\hat{\bold{x}}}$ operator. In the calculation of Fig.~3(c), the truncation order of $\alpha$ is 5 along the $x$-axis and 200 along the $y$-axis: $|n_1|\leq5, |n_2|\leq200$ with $G_{\alpha=(n_1,n_2)}=\frac{2\pi}{L_x}n_1  \hat{\bold{x}} + \frac{2\pi}{L_y}n_2  \hat{\bold{y}}$. We still use the truncation order of $\mu$ of $M=10$ per triangle.

\subsection{1.5. Valley-selective excitation and topological valley transport}
In the presence of an external field $\ket{\phi_{\text{ext}}}$ at a specific frequency $\omega_{\text{ext}}$, the plasmonic equation becomes a linear problem:
\begin{equation}
\ket{\phi}=\ket{\phi_{\text{ext}}}+\left(\omega_{\text{ext}}(\omega_{\text{ext}}+i\gamma)\right)^{-1}\hat{V}_{\bold{k}\rightarrow \bold{0}}\hat{X}_{\bold{k}\rightarrow \bold{0}}\Ket{\phi} \Longrightarrow \ket{\phi}=\left(\bold{I}-\left(\omega_{\text{ext}}(\omega_{\text{ext}}+i\gamma)\right)^{-1}\hat{V}_{\bold{k}\rightarrow \bold{0}}\hat{X}_{\bold{k}\rightarrow \bold{0}}\right)^{-1}\ket{\phi_{\text{ext}}}.
\label{eqs18}
\end{equation}
We calculate $\ket{\phi}_{\text{ext}}$ from the array of point dipoles (see Fig.~\ref{fig.S3.}(a)), considering the MG screening in a similar method described in the section $\bold{1.1.}$. The matrix elements of the basis transformation operator $\hat{B}_{\bold{k}\rightarrow\bold{0}}$ are evaluated using Eq.~(\ref{eqs17}).

As mentioned in the main text, our numerical method intrinsically imposes periodic boundary conditions; thus, as depicted in Fig.~\ref{fig.S3.}(b), we are actually simulating over a 2D periodic array of the dipole arrays. To minimize the effect of the dipole array from a unit cell into the adjacent cells, we use a large domain size ($50\times32=1600$ triangles), and introduce a finite Drude loss ($\omega_{\text{ext}}/\gamma=300$) so that the excited edge mode decays to a negligible amplitude during the propagation within a domain (Fig.~\ref{fig.S3.}(c)). In the calculations of Fig.~4, the truncation order of $\alpha$ is 150 along the $x$-axis and 96 along the $y$-axis; thus, $\hat{V}_{\bold{k}\rightarrow \bold{0}}$ and $\hat{X}_{\bold{k}\rightarrow \bold{0}}$ are $58093\times58093$ complex matrices. In order to reduce the appreciable memory requirement, we lower the truncation order of $\mu$ to 5 per triangle; still, $\hat{B}_{\bold{k}\rightarrow \bold{0}}$ is a complex $64000\times58093$ matrix.

\section{2. Non-scalable Corrections and Landau Damping}
The plasmonic dispersions are not scalable under consideration of non-scalable corrections. In this section, we discuss the significance of these corrections and estimate the lower bound of the system size, above which graphene plasmons don't suffer from additional losses due to intra-/interband electronic transitions.

\subsection{2.1. Quantum capacitance of graphene and non-local effects}
\begin{figure}[t]
    \includegraphics[width=0.8\columnwidth]{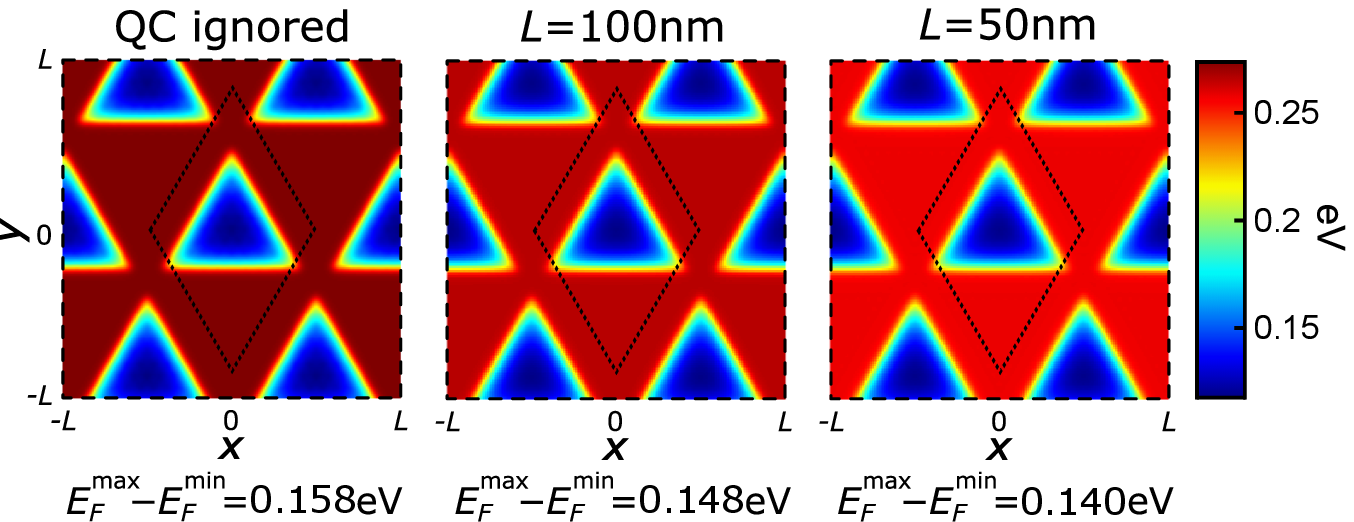}
    \caption{The correction from quantum capacitance (QC) of graphene. The most left calculation ignores QC, and is the same to Fig.~2(b) in the main text. The QC correction reduces the modulation depth of Fermi level, and the effect grows as the system size reduces.}
    \label{fig.S4.}
\end{figure}

In electrochemical equilibrium, the static potential energy $\phi_{ES}(\bold{r})$ and the static charge density on graphene $n_{ES}(\bold{r})$ have to satisfy:
\begin{equation}
\phi_{ES}(\bold{r}) + E_F(\bold{r}) = eV \rightarrow \phi_{ES}(\bold{r}) + \hbar v_F \sqrt{\pi n_{ES}(\bold{r})} = eV,
\label{eqs19}
\end{equation}
where $V$ is the voltage difference applied between graphene and MG. The Fermi energy term can be interpreted as the contribution from an additional capacitance that is intrinsic to graphene, which is called the quantum conductance (QC) of graphene \cite{QCG}. Eq.~(\ref{eqs9}) and Eq.~(\ref{eqs19}) altogether give a self-consistent Thomas-Fermi solution. The difference between this self-consistent solution and the simple electrostatic simulation is negligible when $eV$ is much greater than $E_F(\bold{r})$. If we assume $L/r=4$, $L/h=25$, and $V/h=\SI{1}{\volt/\nano\meter}$, the correction from quantum capacitance of graphene is not very significant unless $L$ goes below $\SI{100}{\nano\meter}$ ($h=\SI{4}{\nano\meter}$), see Fig.~\ref{fig.S4.}.

\begin{figure}[t]
    \includegraphics[width=0.87\columnwidth]{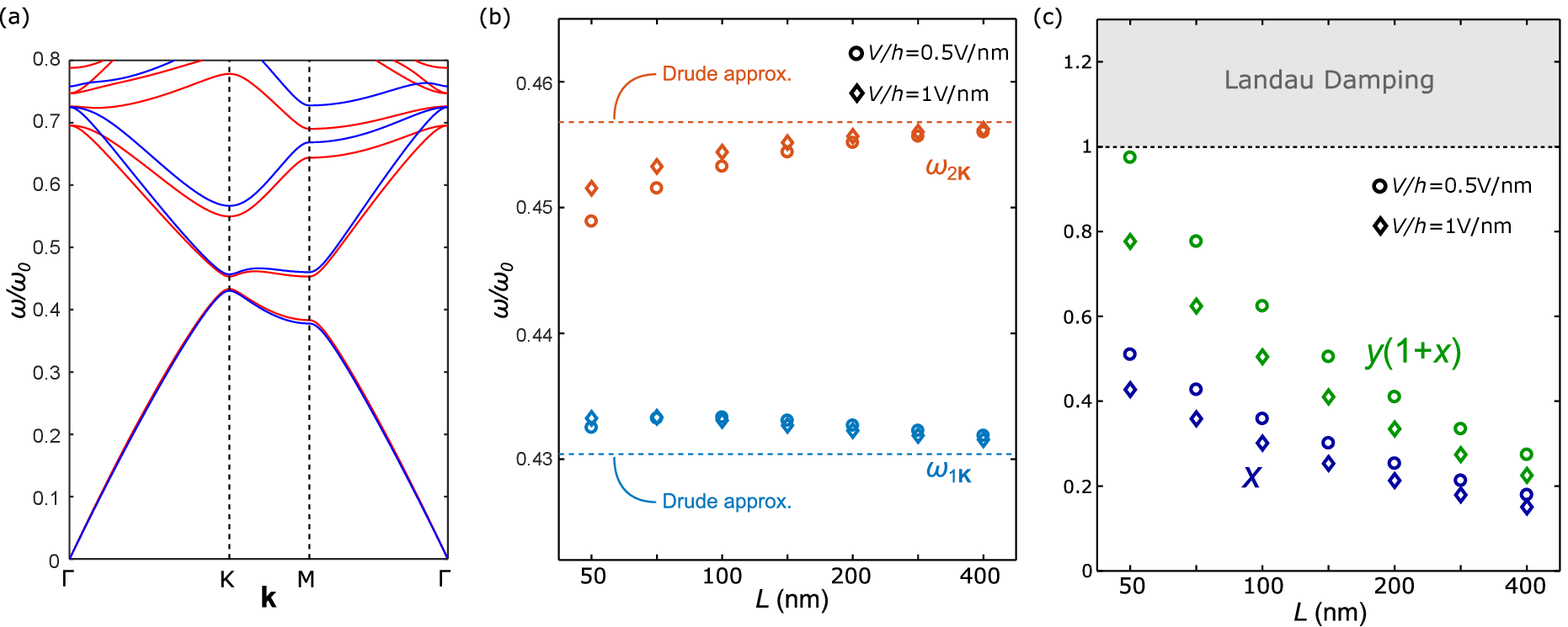}
    \caption{(a) Blue: Band dispersion calculated without non-scalable corrections ($L/r=4$, $L/h=25$, and $\theta=30^\circ$), Red: Band dispersion calculated with non-scalable corrections at $L=100\SI{}{\nano\meter}$ and $V/h=\SI{0.5}{\volt/\nano\meter}$. (b) Non-scalable corrections in $\omega_{1\bold{K}}$ and $\omega_{2\bold{K}}$ ($L/r=4$, $L/h=25$, and $\theta=30^\circ$). (c) Threshold parameters for landau damping ($x$, dark blue) and interband EHPs ($y(1+x)$, brown) evaluated with the mid-gap frequency at $\bold{K}$ and the minimum value of $E_F(\bold{r})$ ($L/r=4$, $L/h=25$, and $\theta=30^\circ$).}
    \label{fig.S5.}
\end{figure}

We calculate the non-scalable corrections in $\omega_{1\bold{K}}$ and $\omega_{2\bold{K}}$ considering both QC of graphene and non-local effects (Eq.~(\ref{eqs15})), see Fig.~\ref{fig.S5.}(b). These corrections reduces the bandgap size and the mid-gap frequency, and their effects decreases as the system size $L$ increases or as the induced Fermi energy $E_F\propto\sqrt{V/h}$ increases. As shown in Fig.~\ref{fig.S5.}(a), the non-scalable corrections are more significant at higher bands.

\subsection{2.2. Intraband and interband transitions}
According to Ref.~\cite{RPA}, even when the Drude loss $\gamma$ is zero, the imaginary part of the Hartree function of 2D massless electrons becomes non-zero under certain energy-momentum ranges:
\begin{subequations}
\label{eqs20}
\begin{gather}
\hbar\omega < \hbar v_F |\bold{q}|\\
\hbar\omega + \hbar v_F |\bold{q}| > 2E_F,
\end{gather}
\end{subequations}
which implies a finite loss of plasmonic excitations regardless of the quality of graphene growth. If we use the dimensionless parameters used in Eq.~(\ref{eqs14}), $x\equiv v_F |\bold{q}|/\omega$ and $y\equiv \hbar \omega / 2E_F$, those conditions are simplied to $x>1$ and $y(1+x)>1$. Actually, these are also the conditions where Eq.~(\ref{eqs15}) is no longer valid. Physically, Eq.~(\ref{eqs20}a), or $x>1$, corresponds to the condition for severe landau damping, and Eq.~(\ref{eqs20}b), or $y(1+x)>1$, corresponds to the condition for the excitation of interband electron-hole pairs (EHP). When the plasmon frequency and momentum lie in those ranges, the plasmonic excitations cannot be sustained even in perfectly clean ($\gamma=0$) graphene. Figure.~\ref{fig.S5.}(b) depicts these threshold parameters, $x$ and $y(1+x)$, at several different system sizes and different doping levels in our system. In the calculations, we take $\omega=\frac{1}{2}(\omega_{1\bold{K}}+\omega_{2\bold{K}})$ and $|\bold{q}|=|\bold{K}|=\frac{4\pi}{3L}$ for rough estimation of those parameters near the topological bandgap. Because the induced Fermi energy is not homogeneous across graphene, we take the minimum value of $E_F(\bold{r})$ for the estimation of $y$. As clearly seen in Fig.~\ref{fig.S5.}(b), our system is far from regions of Landau damping as long as the system size stays reasonable---accessible in the capabilities of up-to-date nanofabrication technologies. Note that $x$ is equal to $\rho_2$ defined in the main text.

\subsection{2.3. Higher order corrections in non-local response and $v_F$-renormalization}
\begin{figure}[t]
    \includegraphics[width=0.7\columnwidth]{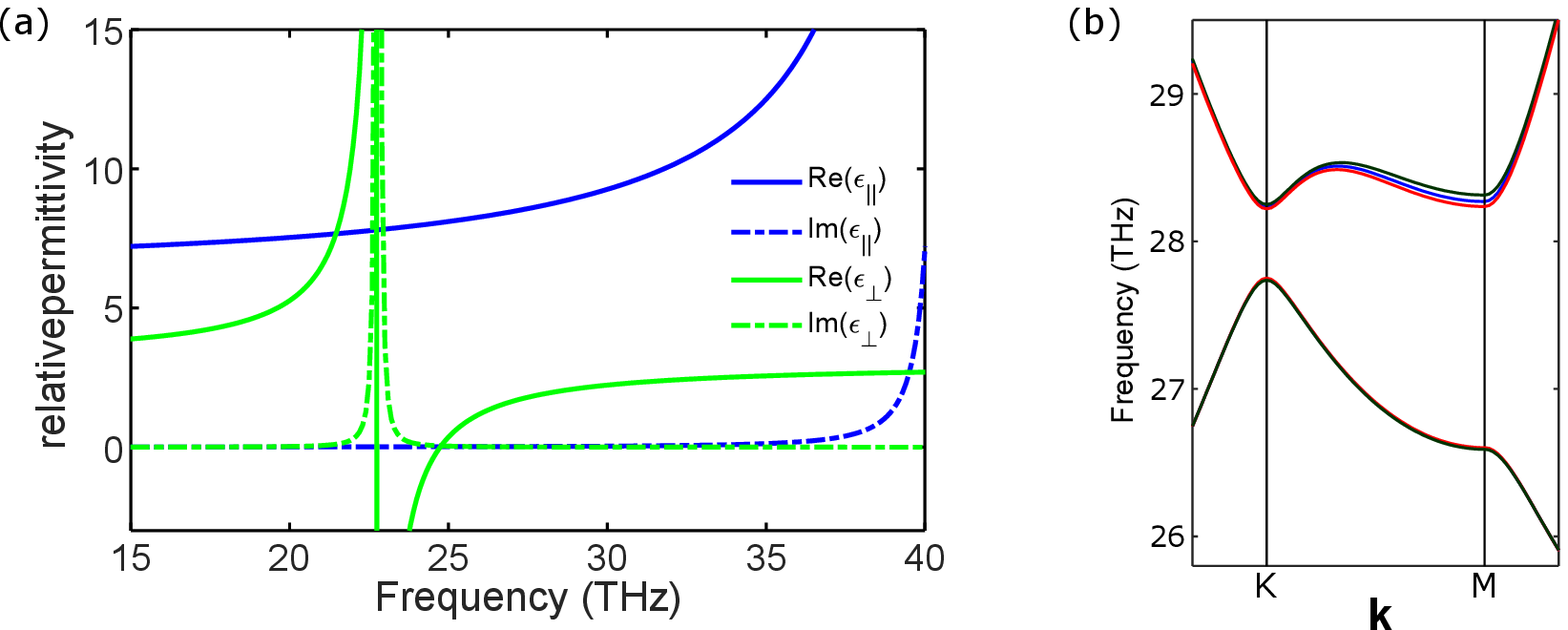}
    \caption{An example of band dispersion in a dispersive medium (hBN) (a) hBN permittivity data from \cite{hBN2, hBN_permittivity} (b) The dark green curve is the dispersion calculated using $\chi_{\text{H}}^{4th}$ (Eq.~\ref{eqs14}) without considering $v_F$-renormalization; the red curve is calculated using $\chi_{\text{H}}^{6th}$ (Eq.~\ref{eqs15}) without considering $v_F$-renormalization; the blue curve is calculated using $\chi_{\text{H}}^{4th}$ with $v_F$-renormalization (Eq.~\ref{eqsvF}) (the graphene quantum capacitance is accounted for all curves)}
    \label{fig.S8.}
\end{figure}

We confirm that it is sufficient to take the terms upto the fourth order in $x$ and $y$ (equivalent to the second order in $|\bold{q}|^2$, as mentioned in the main text) in $\chi_{\text{H}}$ for precise determination of the bandgap size and location; Figure.~\ref{fig.S8.}(b) shows that there is no significant difference between the dark green and the red curve.

We also confirm that the effect of $v_F$-renormalization is negligible. To model $v_F$-renormalization, we use the density-dependent renormalized Fermi velocity expression from Ref.~\cite{novoselov_pnas13}:
\begin{equation}
v_F(n) = v_F(n_0)\left(1 + \frac{1}{4\pi}\frac{\epsilon_{gr}}{\epsilon_{gr}+\epsilon_\text{BN}} \text{ln}(n_0/n) \right),
\label{eqsvF}
\end{equation}
where $n_0=10^{15}$cm$^{-2}$ is the cut-off density, $v_F(n_0)=0.85 \times 10^6$m/s, $\epsilon_{gr} = e^2/8\hbar v_F\epsilon_0 \sim 4$ is the graphene screening, and $\epsilon_\text{BN}\sim3.5$ is the static permittivity of hBN along z-direction. When we ignore $v_F$-renormalization in Fig.5. in the main text, or in the green and red curves in Fig.~\ref{fig.S8.}, we take the value of $v_F$ to be the average value of $v_F(n(\bold{r}))$ over a unit cell, which was $\sim 1.02\times 10^6$m/s. As clearly seen in the negligible difference between the dark green and blue curves in Fig.~\ref{fig.S8.}, the effect of $v_F$-renormalization can be effectively ignored.

\section{3. Hermiticity of the Plasmonic Eigenvalue Equation and Valley-Chern Number}
Here, we confirm that the plasmonic eigenvalue equation given in Eqs.~(2) in the main text is Hermitain under an appropriate inner product. Our discussion given in this section shares similarities to the discussion given in Ref.~\cite{TGP} (see the second section of their supplemental material).
\subsection{3.1. Generalized Hermitian eigenvalue problem}
The plasmonic eigenvalue equation in Eqs.~(2) is $\hat{V}_{\bold{k}}\hat{X}_{\bold{k}}\Ket{\phi_{n\bold{k}}} = \lambda_{n\bold{k}} \Ket{\phi_{n\bold{k}}}$, where $[\hat{V}_{\bold{k}}]_{\alpha, \beta} = v_{\bold{k}}\left(\bold{G}_{\alpha}, \bold{G}_{\beta},\omega_{n\bold{k}}\right)$, $[\hat{X}_{\bold{k}}]_{\alpha, \beta} = \omega^{2} \tilde{\chi}_{nn}\left(\bold{k}+\bold{G}_{\alpha},\bold{k}+\bold{G}_{\beta},\omega_{n\bold{k}}\right)$, $[\Ket{\phi}_{n\bold{k}}]_{\alpha} = \phi_{sc}(\bold{k}+\bold{G}_\alpha,\omega_{n\bold{k}})$, and $\lambda_{n\bold{k}} = \omega_{n\bold{k}}^2$. In previous sections, we observed that both $\hat{V}_{\bold{k}}$ and $\hat{X}_{\bold{k}}$ are Hermitians; however, their product is not Hermitian under the standard inner product.
Nevertheless, we can simply recast this problem as a generalized eigenproblem by multiplying the both sides by the inverse of the Coulomb operator $\hat{V}_{\bold{k}}^{-1}$:
\begin{equation}
\hat{X}_{\bold{k}}\Ket{\phi_{n\bold{k}}} = \lambda_{n\bold{k}} \hat{V}_{\bold{k}}^{-1}\Ket{\phi_{n\bold{k}}}.
\label{eqs21}
\end{equation}
Then, Eq.~(\ref{eqs21}) is a generalized eigenvalue equation with Hermitian definite pencil $\{\hat{X}_{\bold{k}},\hat{V}_{\bold{k}}^{-1}\}$ \cite{Her}.

\subsection{3.2. Berry curvature and valley-Chern number}
In the calculations of Berry curvature, we use the method described in Ref.~\cite{Fukui} is. The detailed description of the method can be also found in the supplemental material of Ref.~\cite{TGP}, Here, we briefly note two important comments regarding the calculations of Berry curvature and valley-Chern numbers. First, when we evaluate an inner product between two state vectors, let's say $\ket{\phi}$ and $\ket{\phi'}$, during the calculations of Berry curvature, their inner product is defined respect to the Hermitian pencil $\{\hat{X}_{\bold{k}},\hat{V}_{\bold{k}}^{-1}\}$:
\begin{equation}
\braket{\phi|\phi'} \equiv \sum_{\alpha, \beta} \left[\hat{V}_{\bold{k}}^{-1}\right]_{\alpha, \beta} \left[\ket{\phi}\right]_\alpha^{*} \left[\ket{\phi'}\right]_\beta.
\label{eqs22}
\end{equation}

Second, as mentioned in the main text, we define the valley-Chern number $C_\nu$ to be the integration of the berry curvature over a triangular area centered at the valley, where the vertices of the triangle are given as the nearest $\Gamma$ points from the valley. We come up with this artificial definition because two valleys actually share the same momentum space in contrary to the assumption considered in the calculation of $C_\nu$ using effective Hamiltonians \cite{VHI2007} that two valleys are separated into two toally decoupled subspaces. The triangular area defined above is the largest momentum subspace, in which the valley of our interest in $C_\nu$ calculation is the closest, and occupies exactly the half of the whole Brillouine zone. The calculated values of $|C_\nu|$ in Fig.~3(a) in the main text show small deviations from 0.5 for three reasons: (1) our choice of the integration range cannot exclude the contributions from $\Gamma$-points that are not relevant to valley-based dynamics, (2) our system is not a two-level system, and (3) the contributions from the adjacent valleys are still present because the valleys share the same momentum space.

\begin{figure}[t]
    \includegraphics[width=0.8\columnwidth]{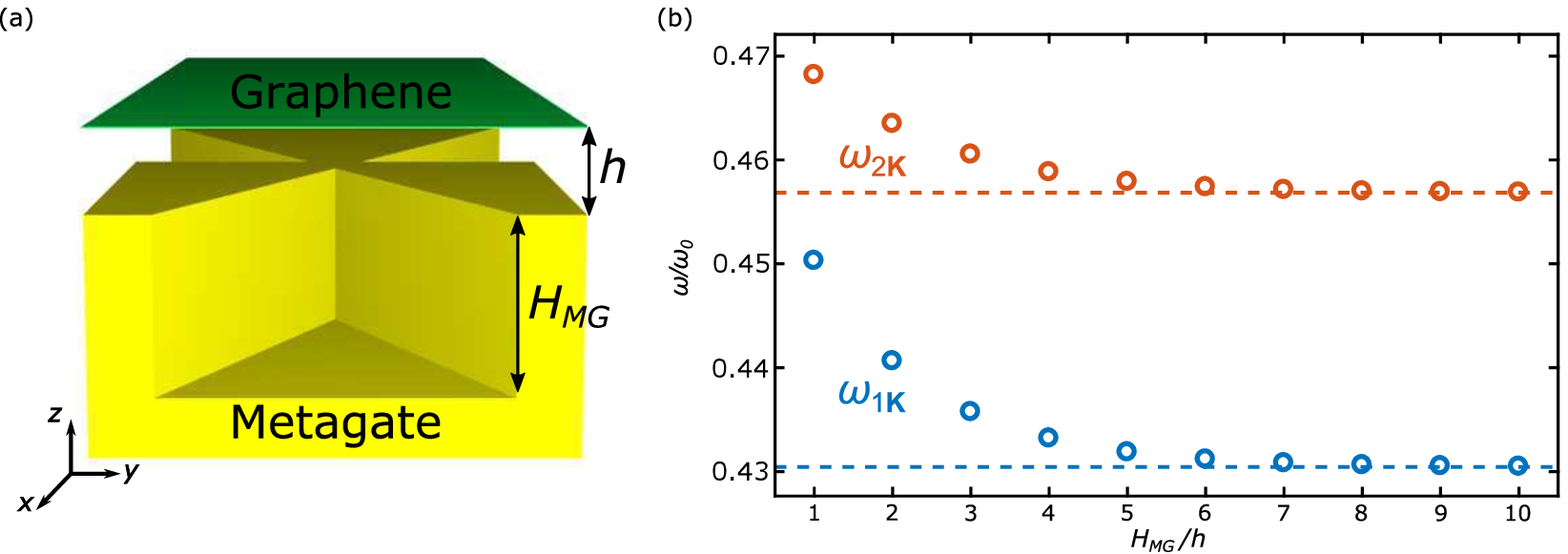}
    \caption{Effect of the finite depth of the triangular holes in MG onto the topological bandgap (a) A realistic MG structure achievable via etching; side view. (b) Shifts in the plasmonic frequencies $\omega_{1,2\bold{K}}$ as a function of the depth of the holes ($L/r=4$, $L/h=25$, $\theta=30^\circ$); the dotted lines indicates the calculated frequencies under the assumption $H_{\text{MG}}\rightarrow\infty$.}
    \label{fig.S6.}
\end{figure}

\section{4. Possible Perturbations on Metagate structure}
So far, we have assumed that the holes in MG are infinitely deep and perfectly triangular. In this section, however, we break these two assumptions, thereby perturbing the geometry of the holes in MG. Then, we examine how much the size of the topological bandgap is robust against possible perturbations on the structure of MG that may occur due to imperfection in nanofabrication.

\begin{figure}[b]
    \includegraphics[width=0.8\columnwidth]{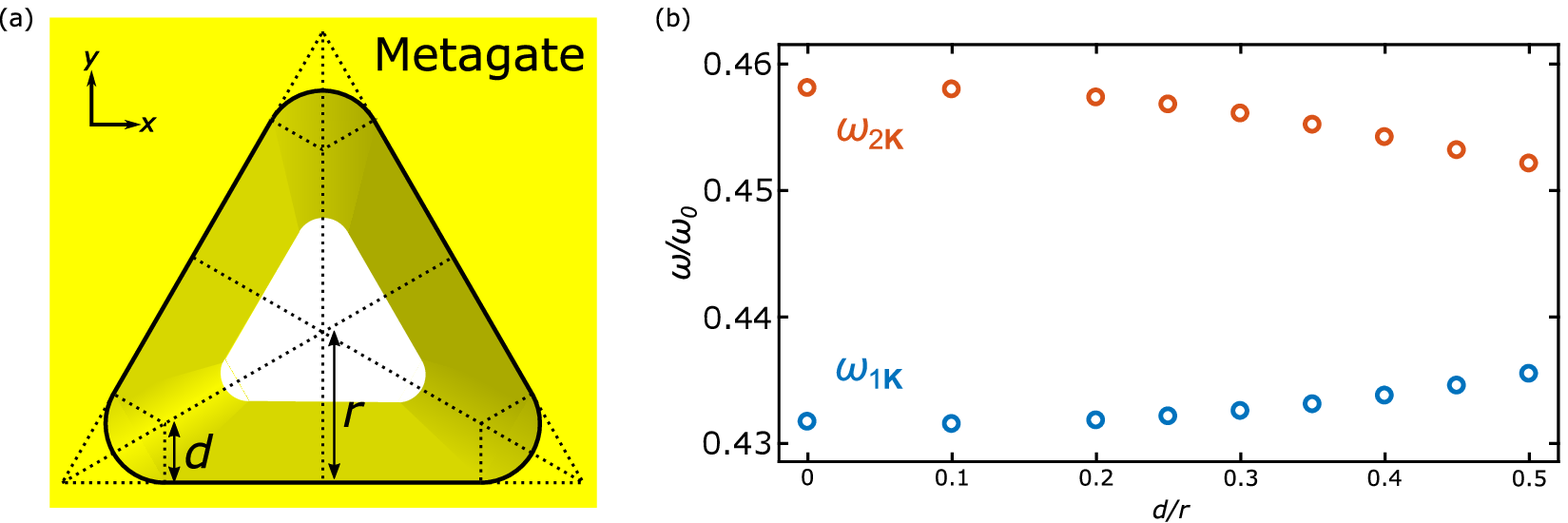}
    \caption{Effect of the finite radius of curvature at the corners of the triangular holes (a) A realistic MG structure achievable via etching following after electron-beam lithography; top view. (b) Shifts in the plasmonic frequencies $\omega_{1,2\bold{K}}$ as a function of the radius of curvature at the corners of the triangle ($L/r=4$, $L/h=25$, $\theta=30^\circ$).}
    \label{fig.S7.}
\end{figure}
\subsection{4.1. Finite aspect ratio of the triangular holes}
First, we evaluate the effect of finite depth of the holes in MG. We consider a structure depicted in Fig.~\ref{fig.S6.}(a) that the other end of the holes are terminated by the same metallic material. In order to take the finite depth of the holes into account in our numerical model, we simply modify the ansatz for the potential in the holes $\phi_{\text{hole}}(\bold{r},z)$ in Eq.~(\ref{eqs5}) so that $\phi_{\text{hole}}(\bold{r},z=-h-H_{\text{MG}})=0$:
\begin{equation}
\begin{split}
\phi_{\text{hole}}\left(\bold{r},z\right) = \sum_{\mu}t_{\mu}T_{\mu}\left(\bold{r}\right)\left(\cosh\left(\xi_{\mu}(z+h)\right)+\coth\left(\xi_{\mu}H_{\text{MG}}\right)\sinh\left(\xi_{\mu}(z+h)\right)\right),
\end{split}
\label{eqs23}
\end{equation}
where $H_{\text{MG}}$ is the depth of the holes. Then, the matrix elements of $\hat{\Xi}$ needs to be redefined accordingly for the third boundary condition matching in Eq.~(\ref{eqs6}): $[\hat{\Xi}]_{\mu, \nu}=\xi_\mu \coth\left(\xi_{\mu}H_{\text{MG}}\right) \delta_{\mu\nu}$. The rest of the discussion in the section $\bold{1.1.}$ following after Eq.~(\ref{eqs6}) applies in the same way.

Figure.~\ref{fig.S6.}(b) illustrates that the topological bandgap at $\bold{K}$ point is highly insensitive to $H_{\text{MG}}$ as long as it is greater than $\sim$5 times $h$. This is due to the ultrahigh field confinement of acoustic plasmons in graphene with metal \cite{hBN1}. Given that $L/r=4$, $L/h=25$, and $H_{\text{MG}}=5h$, the required aspect ratio of the triangular holes $2\sqrt{3}r/H_{\text{MG}}$ (the ratio of the side length of the triangle to the height of the holes) is estimated to be around $4$. For example, if $L$ is $\SI{200}{\nano\meter}$, the side length of the triangle is $\SI{173}{\nano\meter}$, and the required height of the holes is $\SI{40}{\nano\meter}$ (assuming $H_{\text{MG}}=5h$). Thus, the assumption that the holes are infinitely deep works well as long as the triangular shape of the holes are maintained for a relatively shallow depth.

\subsection{4.2. Round corners of the triangle}
Next, we introduce a finite radius of curvature at the corners of the triangle as illustrated in Fig.~\ref{fig.S7.}(a). In calculation of Fig.~\ref{fig.S7.}(b), we choose to use COMSOL Multiphysics Electromagnetic Waves, Frequency Domain module instead of our in-house codes since it is tricky to compute the matrix elements of $\hat{B}_{\bold{k}}$ when the analytic expression for the Laplacian eigenfunctions is not known. Assuming $L=\SI{200}{\nano\meter}$ and $r=\SI{50}{\nano\meter}$, the reduction in the size of the topological bandgap, $\omega_{2\bold{K}}-\omega_{1\bold{K}}$, is calculated to be only around $3\%$ with $d=\SI{10}{\nano\meter}$ ($d/r=0.2$). Thus, our theoretical predictions made with the assumption that the holes are perfectly triangular are still sufficiently precise despite roundy corners of the triangle in real samples, given that a resolution of $d=\SI{10}{\nano\meter}$ (the minimum diameter of a dot is $\SI{20}{\nano\meter}$) is well within experimental capabilities in electron-beam lithography \cite{ebeam} or focused ion-beams.

\end{document}